\documentclass[aps,prb,twocolumn,showpacs,floatfix,a4paper]{revtex4}
\usepackage{graphicx}
\usepackage[LGR,T1]{fontenc}
\usepackage{geometry}
\usepackage{units}
\usepackage{amsmath}
\usepackage{amssymb}
\usepackage{color}

\usepackage{esint}
\usepackage{bm}
\usepackage{ulem}
\usepackage[colorlinks=true, citecolor=blue]{hyperref}
\setcitestyle{journalcolor=blue}

\usepackage{tikz}
\usetikzlibrary {arrows.meta, bending}
\newcommand{\rungR}[4]
{
	\draw[black] (#1, #2) to (#1, #3);
	\filldraw (#1, #2) circle (#4);
	\filldraw (#1, #3) circle (#4);
}
\newcommand{\rungD}[3]
{
	\node[anchor=north] at (#1,#2+0.3) {$\uparrow\downarrow$};
	\draw (#1, #2) circle (#3);

}
\newcommand{\rungB}[3]
{
	\filldraw (#1, #2) circle (#3);

}

\newcommand{\bcol} {\textcolor{blue}}

\newcommand{\ket}[1] {{\left|#1\right\rangle}}

\begin{document}

	\title{Quantum Phases of a Strongly Disordered Two-Legged Josephson Ladder}

	\author{Eyal Walach}
	\author{Efrat Shimshoni}

	\affiliation{Department of Physics, Jack and Pearl Resnick Institute and the Institute of Nanotechnology and Advanced Materials,
		Bar-Ilan University, Ramat-Gan 52900, Israel}

\date{\today}

\begin{abstract}
Disordered superconductors in low dimensions provide an exemplary manifestation for the role of quantum fluctuations in a many-body system. Specifically in Josephson arrays with comparable Josephson and charging energies ($E_J\sim E_C$), disorder tends to change the nature of the paradigmatic Superconductor-Insulator Transition (SIT) and potentially leads to formation of multiple distinct phases. We address this problem 
in a model of a two-legged Josephson ladder subjected to a wide spatial distribution of its parameters along the legs. In contrast, we assume the system to have a perfect $\mathbb{Z}_2$ symmetry to interchange between the legs, and investigate the effects of spatial randomness which preserves this symmetry in the strong-disorder limit. To this end, we apply a strong randomness real-space renormalization group technique and explore the resulting phase diagram. We identify three disorder-dominated phases, including an intermediate phase between a disordered superconductor and a disordered insulator. The latter insulating phase can be mapped to a XY spin-chain in a spin glass phase, while the intermediate phase turns out to be a Bose glass.

\end{abstract}

	\maketitle
	
\section{Introduction and Principal Results}
\label{sec:Introduction}

Superconducting materials and devices provide fertile grounds for observation of quantum behaviors. Cooper pairs take part in a macroscopic wavefunction, hence sharing a global phase. When restricted to low dimensions, quantum fluctuations are enhanced\cite{Hebard1994, RevModPhys.69.315, 10.1063/1.882069, Raychaudhuri_2022,BashanTulipman2025} and the full-fledged superconductivity turns fragile; as a result, the electronic state tends to break into small islands coupled by Josephson interactions. In such systems, repulsive Coulomb interactions lead to behaviors affected by the number-phase uncertainty. When the charging energy $E_C$ and the Josephson energy $E_J$ are of the same order of magnitude, these strongly quantum behaviors give rise to manifestations of macroscopic quantum behaviors; their pioneering demonstration was recently awarded the 2025 Nobel Prize in Physics\cite{PhysRevLett.55.1908}.

When tuning the ratio of these two energy scales, along with other parameters, the system undergoes a superconductor-insulator transition (SIT), a dramatic manifestation of these quantum fluctuations. Its clearest signature is a drastic change in the electric resistance as $T \to 0$; the resistance is zero when the system is superconducting (SC), and it switches to infinity upon tuning of some non-thermal parameter, such as the layer thickness, gating, external magnetic field or various disorder types, beyond a critical value\cite{Hebard1994, RevModPhys.69.315, 10.1063/1.882069}.
This phenomenon is a prototypical example of a Quantum Phase Transition (QPT)\cite{Sachdev_2011}: a fundamental change in the nature of the ground state achieved while maintaining $T=0$.

The interplay of charging and Josephson energies in effectively two-dimensional (2D) or one-dimensional (1D)  superconductors was demonstrated to be captured by models of Josephson arrays or, equivalently, interacting Bosons\cite{PhysRevB.40.546,  PhysRevB.39.2756, PhysRevLett.65.923, PhysRevLett.64.587, PhysRevLett.69.828, PhysRevB.48.3316, PhysRevB.49.9794, PhysRevLett.92.015703,   PhysRevB.94.134501}. For Josephson arrays, the ratio $E_C/E_J$ plays a crucial role -- the SIT occurs at a critical value where $E_C/E_J \sim 1$, where the system has maximal phase-charge uncertainty and quantum fluctuations are most notable.

Realistic material platforms available for observation of the SIT (amorphous metallic films, 2D materials and engineered Josephson arrays) are typically introduced with disorder of various types. These include spatial fluctuations in $E_C/E_J$ (e.g. due to emergent inhomogeneity in superfluid stiffness\cite{Trivedi_2012, PhysRevB.89.035149}), random gating voltage and random magnetic fields. The interplay between various disorder types and interactions may introduce multiple quantum critical points separating more than two phases\cite{RevModPhys.91.011002}. Indeed, different studies have suggested various insulating phases in such systems, including a "Bose/Mott glass" \cite{PhysRevB.40.546, PhysRevB.94.134501, PhysRevB.112.134510}. Some works have shown indications for a more intriguing option -- systems with more than two phases, where an intermediate metallic phase separates the insulating phase and the superconducting phase \cite{Phillips_Science302, RevModPhys.91.011002, PhysRevB.93.205116, PhysRevB.96.245140, NPJ.LiKivelsonLee, ZhangPalevskiKapitulnikPNAS}. Experimental works also show evidence for more complicated phase diagrams in low-dimenstional systems\cite{Chen2021, PhysRevB.110.L180502}. 
The strong fluctuations in the vicinity of the SIT might also introduce a mixed phase where both SC and charge-density correlations form, especially in the presence of disorder.

In a previous work, we have investigated the role of disorder in a quasi-one-dimensional system: a two-legged Josephson ladder\cite{PhysRevB.103.245301}. The clean version of this system is known\cite{PhysRevB.83.220518} to support a superconducting phase, a Mott insulating phase, and an intermediate phase. We have shown that in the weak-disorder limit this system develops, in the strongly fluctuating regime, disorder-dominated phases, including one phase that appeared to be intermediate between a SC and insulator. The perturbative renormalization group (RG) analysis has shown that some disorder types are relevant in various phases, hence the weak-disorder approximation used there does not hold. In this work we apply the opposite approximation and assume that the disorder is very strong.

Generally speaking, there are different possible effects of disorder on such intermediate phase. On one hand, the intermediate phase might appear to be unstable, and disappear due to the effects of disorder. On the other hand, disorder supports inhomogeneity which might \textit{stabilize} intermediate behaviors. Introducing different types of disorder and using strong-disorder approximations we can address these competing effects and explore the nature of the disorder-dominated phases. 

Notably, unlike the clean version of the two-legged Josephson ladder model which can be split into two independent sectors, one Bosonic and one Fermionic, some types of disorder couple these sectors. Such coupling terms were shown to profoundly change the critical behavior of the combined system\cite{PhysRevLett.102.176404, PhysRevLett.114.090404, PhysRevB.95.075132}. In Ref. \onlinecite{PhysRevB.103.245301}, we showed that in the perturbative limit this leads to formation of a rich phase diagram. The present study indicates that the strongly-disordered system maintains certain features of this phase diagram, yet provides a more detailed insight on the nature of the disorder-dominated phases and the transitions between them.

The Josephson ladder investigated in this work assumes strong spatial fluctuations along the legs, so that both intra- and inter-leg charging interactions and Josephson energies are random variables characterized by broad distributions.  
At the same time, a perfect $\mathbb{Z}_2$ symmetry relates the legs at an arbitrary longitudinal site. This symmetry is a crucial feature of our model; 
it is what allows for a phase diagram more complex than the one obtained in a 1D chain with no additional symmetry, as was investigated in previous works\cite{PhysRevLett.93.150402, PhysRevLett.100.170402}. As we show below, certain features characterizing the system behavior are reminiscent of other strongly-disordered 1D systems 
such as random spin models\cite{PhysRevB.36.536, PhysRevLett.69.534, PhysRevB.51.6411}.

Our analysis of the random Josephson ladder model utilizes a real-space strong-disorder renormalization-group (SDRG) method, which provides access to parameter regimes not included in the weak-disorder analysis\cite{PhysRevB.103.245301}. Following \onlinecite{PhysRevB.81.174528} we iteratively decimate high-energy degrees of freedom, leading to an effective Hamiltonian of progressively shorter chains. Alongside a numerical calculation implementing this iterative procedure, we derive master equations for the parameter distribution functions, which provide analytical solutions in certain limit cases. Combining these approaches allows us to identify the stable phases of the system. In particular, 
a phase diagram consisting of three distinct phases is apparent, indicating that the intermediate phase reported in Ref. \onlinecite{PhysRevB.103.245301} exists even as disorder grows strong.

Our main results are summarized in the phase diagram depicted in Fig. \ref{fig:heatmap}, which exhibits   three stable phases.
For strong Josephson couplings and relatively modest charging energies (rightmost side of the phase diagram) one obtains a disordered superconductor. The system undergoes a Kosterlitz-Thouless like transition\cite{KosterlitzThouless_1973, PhysRevB.81.174528} when charging energy is increased, turning into an intermediate phase of a Bose-glass character: it establishes phase-coherence on the rungs but manifests a suppressed charge transport between adjacent rungs. This behavior allows for a superfluid transport of anti-symmetric (quasi-spin) current, flowing in opposite directions on the different legs, but insulating behavior 
for longitudinal charge current. When charging energy is further increased and Josephson energy decreases, the system is mapped to a phase dominated by spin-like degrees of freedom where every rung is half-filled, i.e. hosting exactly one charge per the  two sites; this yields segments of spin chains in a spin glass phase which dominate the system's behavior.


The structure of the paper is as follows. In Sec. \ref{sec:Model} we define the ladder model and briefly discuss physical realizations. In Sec. \ref{sec:RGFlow} we introduce the basic possible RG steps (\ref{subsec:Elementary}), discuss the Master Equations for the distributions of parameters (\ref{subsec:MasterEqns}) and pose an ansatz for these distributions (\ref{subsec:Ansatz}); we then elaborate on details and limitations of the numerical simulation (\ref{subsec:SimulationDetails} and \ref{subsec:SpinChain}). In Sec. \ref{sec:Results} we discuss the numerical results, compare them to analytical solutions of the RG flow equations and the posed ansatz, and draw conclusions regarding the resulting phases and phase transitions. In Sec. \ref{sec:Analysis} we discuss the results and the nature of the phases, summarize the main results and provide an outlook. Finally, in the appendices we give the details of our calculations.

\section{Model}
\label{sec:Model}

Our starting point for the theoretical analysis is a superconducting two-legged ladder, composed of phase-coherent grains which lie on two parallel legs connected by rungs 
as depicted in Fig. \ref{fig:Ladder}. Each grain is characterized by a number of Cooper pairs $n_{i,\nu}$ and a phase $\phi_{i,\nu}$ of the superconducting order parameter (with $\nu=u,d$ the leg index and $i$ the site on the leg), which obey the canonical commutation relations $[\phi_{i,\nu},n_{i',\nu'}]=\textrm{i}\delta_{i,i'}\delta_{\nu,\nu'}$. The Hamiltonian can be written as follows:

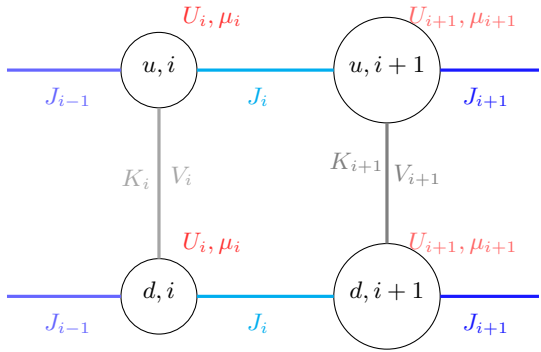
\begin{figure}
\begin{tikzpicture}
    \draw (0, 0) circle (0.5) node[left=0, above=-0.2cm] {$u,i$};
    \draw (0, -3) circle (0.5) node[left=0, above=-0.2cm] {$d,i$};
    \node[red!80] at (0.7, 0.7) {$U_i, \mu_i$};
    \node[red!80] at (0.7, -2.3) {$U_i, \mu_i$};
    \draw[very thick, gray!70] (0, -0.5) to (0, -2.5) node[left=0.3cm, above=0.8cm] {$K_i$};
    \node[gray!70] at (0.3, -1.4) {$V_i$};

    \draw (3, 0) circle (0.7) node[left=0, above=-0.2cm] {$u,i+1$};
    \draw (3, -3) circle (0.7) node[left=0, above=-0.2cm] {$d,i+1$};
    \node[red!60] at (4.0, 0.7) {$U_{i+1}, \mu_{i+1}$};
    \node[red!60] at (4.0, -2.3) {$U_{i+1}, \mu_{i+1}$};
    \draw[very thick, gray!100] (3, -0.7) to (3, -2.3) node[left=0.4cm, above=0.8cm] {$K_{i+1}$};
    \node[gray!100] at (3.4, -1.4) {$V_{i+1}$};

    \draw[very thick, blue!60] (-2, 0) to (-0.5, 0) node[left=0.7cm, below=0.1cm] {$J_{i-1}$};
    \draw[very thick, blue!60] (-2, -3) to (-0.5, -3) node[left=0.7cm, below=0.1cm] {$J_{i-1}$};

    \draw[very thick, cyan] (0.5, 0) to (2.3, 0) node[left=1.0cm, below=0.1cm] {$J_i$};
    \draw[very thick, cyan] (0.5, -3) to (2.3, -3) node[left=1.0cm, below=0.1cm] {$J_i$};

    \draw[very thick, blue!90] (3.7, 0) to (5, 0) node[left=0.7cm, below=0.1cm] {$J_{i+1}$};
    \draw[very thick, blue!90] (3.7, -3) to (5, -3) node[left=0.7cm, below=0.1cm] {$J_{i+1}$};
\end{tikzpicture}
\caption{\label{fig:Ladder}
The basic system described in equations (\ref{eqn:Hsum})-(\ref{eqn:Hrung}).  Different colors correspond to different Hamiltonian terms, and different hues stand for different energies (either Josephson for the bonds or Charging for the sites). Note that while every bond and every site have different energies, the up-down symmetry is preserved.\\
}
\end{figure}

\begin{align}
\mathcal{H} & = \sum_i \left[\mathcal{H}_{u,i}+\mathcal{H}_{d,i}+\mathcal{H}_{int,i}\right] \label{eqn:Hsum}\\
\mathcal{H}_{\nu,i} & = \left[ \frac{1}{2} U_i n_{\nu i}^2 - \mu_i n_{\nu i} - J_i \cos(\phi_{\nu , i} - \phi_{\nu ,i+1})\right]\label{eqn:Hleg}\\
\mathcal{H}_{int,i} & =\left[-K_i\cos\left(\phi_{u, i}-\phi_{d, i}-\varphi_{K i}\right)+V_in_{u,i}n_{d,i}\right].\label{eqn:Hrung}
\end{align}
Here $U$ is the charging energy on the leg; $J$ is the intra-leg Josephson coupling; $K e^{i\varphi_K}$ is the inter-leg Josephson coupling, where the phase $\varphi_K$ accounts for a contribution from, e.g., a magnetic flux threading the plane of the ladder; $V$ is the inter-leg charging energy, and $\mu$ is the chemical potential. Note that $U, J, \mu$ are random variables which depend on $i$ but not on $\nu$: we assume that the disorder respects the $\mathbb{Z}_2$ up-down symmetry between the legs.

For every site, we will denote
\begin{align}
    \bar{n}_i = \frac{\mu_i}{U_i+V_i} \label{eqn:nbar}
\end{align}
or the "average filling" to be the number such that $n_u=n_d=\bar{n}$ minimizes the charge sector of the energy. It will turn out that the value of $\bar{n}$ changes the ground state of the charge sector in a non-trivial way; this will be discussed in \ref{subsec:Elementary}.

Note that $\bar{n}_i$ can be defined modulo $1$, and we will assume that $\left|\bar{n}_i\right|\leq \frac{1}{2}$.
Also note that $V>U$ leads to an unstable behavior, hence we assume $V<U$; we further focus on the repulsive interaction regime $0<V$ as it was shown to give rise to more interesting behaviors\cite{PhysRevB.103.245301}.

As mentioned in the introduction, the perfect $\mathbb{Z}_2$ symmetry of the model is crucial -- otherwise, it will behave like a one-dimensional system with next-nearest-neighbors interactions and hopping. The leg index, $\nu$, is therefore not likely to represent physical separation, but rather some kind of pseudo-spin degree of freedom.


\section{Derivation of the RG Flow Equations}
\label{sec:RGFlow}

Following Refs. \onlinecite{PhysRevLett.93.150402, PhysRevLett.100.170402,
PhysRevB.81.174528} we construct a strong-disorder renormalization-group scheme (SDRG). The leading idea of SDRG is to iteratively focus on locally dominant energy scales from the Hamiltonian, assume that the disorder is strong enough so that this energy scale dominates its neighbors, and find an effective description with the related degree of freedom "frozen" to the ground state. Two sites connected by the strongest bond will turn into a single phase-coherent cluster. Similarly, in a site with strong charging energy the particle number $n$ will be locked at its ground-state value, while excited number states will be eliminated. The RG flow is parametrized by the largest energy scale existing in the system, $\Omega$, and we iteratively "decimate" all degrees of freedom with $E>e^{-d\Gamma} \Omega$.

Notably, in this scheme the structure of the Hamiltonian in Equations (\ref{eqn:Hsum})-(\ref{eqn:Hrung}) is not preserved under the RG flow. We call rungs that maintain this structure "regular" sites (abbreviated as "R"), and will soon introduce two other types of sites - "bowties" ("B") and "doublets" ("D"). After a long sequence of RG steps our ladder might include these three types of sites connected in different ways, as can be seen in the unit cells depicted in Figs. \ref{fig:RBRungs} and \ref{fig:DRungs}.

In the following subsections we outline the derivation of RG flow for this system. We start by describing the allowed RG steps for the original rungs, and the origin of the Bowties and the Doublets. We will then derive a master equation formulation and give a detailed ansatz that will lead us to analytical observations about the system. In the results section we will use both these analytical results and numerical simulations based on the RG steps described below.

\subsection{Elementary Steps}
\label{subsec:Elementary}
\begin{figure}
\begin{tikzpicture}
\rungR{0}{0}{-2}{0.1}
\rungR{2}{0}{-2}{0.1}
\draw[blue, very thick] (0, 0) to (2, 0) node[left=1cm, above=0cm] {$J$};
\draw[blue, very thick] (0, -2) to (2, -2) node[left=1cm, above=0cm] {$J$};

\rungR{2.7}{0}{-2}{0.1}
\rungR{4.7}{0}{-2}{0.1}
\draw (2.7, 0) to (4.7, 0);
\draw (2.7, -2) to (4.7, -2);
\draw[cyan, very thick] (2.7, 0) to (4.7, -2) node[left=0.6cm, above=0.7cm] {$J_\times$};
\draw[cyan, very thick] (2.7, -2) to (4.7, 0);

\draw[teal, very thick] (5.4, 0) to (5.4, -2) node[right=0.2cm, above=0.8cm] {$K$};
\fill (5.4, 0) circle (0.1);
\fill (5.4, -2) circle (0.1);
\rungB{7.4}{-1}{0.1}
\draw (5.4, 0) to (7.4, -1) to (5.4, -2);

\rungB{1}{-4}{0.1}
\draw (3, -3) to (3, -5);
\draw (3, -3) to (1, -4) to (3, -5);
\fill[magenta] (3, -3) circle (0.15) node [right=0.5cm, above=0.1cm] {$U, V, \mu$};
\fill[magenta] (3, -5) circle (0.15);

\rungB{4}{-4}{0.1}
\draw (4, -4) to (6, -4);
\fill[purple] (6, -4) circle (0.15) node[right=0.4cm, above=0.1cm] {$U^B, \mu$};


\draw[blue, thick] (0, -6) to (1.9, -6);
\draw[blue, thick] (0, -8) to (1.9, -8);
\draw[teal, thick] (0, -6) to (0, -8);
\fill[magenta, thick] (0, -6) circle (0.1);
\fill[magenta, thick] (0, -8) circle (0.1);
\draw[blue, thick] (1.9, -6) to (3.8, -6);
\draw[blue, thick] (1.9, -8) to (3.8, -8);
\draw[cyan, thick] (1.9, -6) to (3.8, -8);
\draw[cyan, thick] (1.9, -8) to (3.8, -6);
\draw[teal, thick] (1.9, -6) to (1.9, -8);
\fill[magenta, thick] (1.9, -6) circle (0.1);
\fill[magenta, thick] (1.9, -8) circle (0.1);
\draw[blue, thick] (3.8, -6) to (5.6, -7) to (3.8, -8);
\draw[teal, thick] (3.8, -6) to (3.8, -8);
\fill[magenta, thick] (3.8, -6) circle (0.1);
\fill[magenta, thick] (3.8, -8) circle (0.1);
\draw[blue, thick] (5.6, -7) to (7.4, -7);
\fill[purple, thick] (5.6, -7) circle (0.1);
\fill[purple, thick] (7.4, -7) circle (0.1);

\end{tikzpicture}
\caption{\label{fig:RBRungs}
The five unit cells involving R sites and B sites, and a section from a ladder. \\ Top row (left to right): $R_{r=}, R_{r\times}, R_b$. Middle row (left to right): $B_r, B_b$. Bottom row: a connected chain of the four unit cells $R_{r=}$, $R_{r\times}$, $R_b$, $B_b$.
A vertical line denotes $K \cos\left(\phi_u-\phi_d-\varphi_K\right)$; a horizontal or diagonal line are $J \cos\left(\phi_1-\phi_2\right)$; a bold dot is a superconducting grain.
} 
\end{figure}
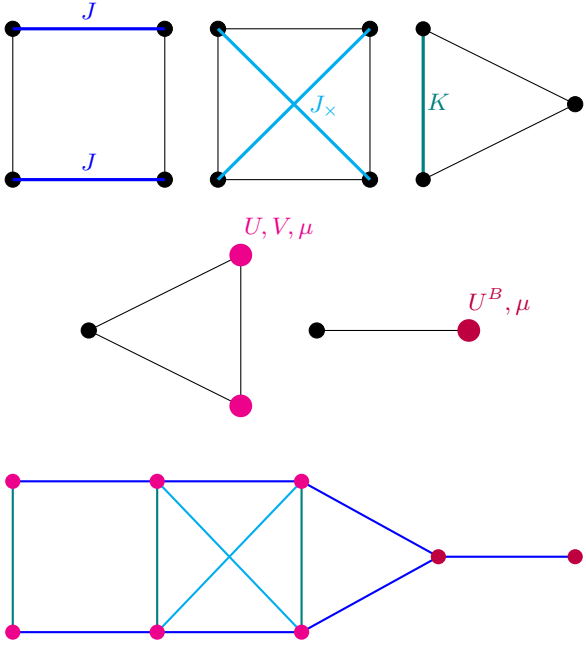

We shall start by explaining the different RG steps that we perform when considering only R sites, and then go on to the possible steps for the other types. The unit cells making the ladder contain not only a site $i$ but also its connections to the neighboring sites. We refer to a unit cell connecting a site of type $X$ to its neighbor of type $Y$ by $X_y$, as can be seen in Fig. \ref{fig:RBRungs}. In several cases, we add an additional symbol in the subindex to differentiate between different bond types -- e.g. $R_{r=}$ and $R_{r\times}$.

Restricting ourselves to R sites, there are three types of decimation steps: inter-site bond decimation (when $J$ is strong), on-site phase-locking ($K$), and on-site charge-locking ($U$).

In a bond decimation step, two adjacent sites unite into one to form two phase-coherent clusters, one on each leg. To find the effective $U, V, \mu$ of the merged clusters we derive the Lagrangian of the rungs. The detailed analysis is shown in appendix \ref{app:Bonds}; here we present the main result. Denoting the sites with the indices $1,2$, the effective parameters are:
\begin{align}
    &K_{eff} e^{i\varphi_{K_{eff}}} = K_{1} e^{i\varphi_{K_{1}}} + K_{2} e^{i\varphi_{K_{2}}} \label{eqn:Jdecim1} \\
    &\bar{n}_{eff} = \bar{n}_1 + \bar{n}_2 \mod 1 \label{eqn:Jdecim2} \\
    &\frac{1}{U_{eff}+V_{eff}} = \frac{1}{U_{1}+V_{1}} + \frac{1}{U_{2}+V_{2}} \label{eqn:Jdecim3} \\
    &\frac{1}{U_{eff}-V_{eff}} = \frac{1}{U_{1}-V_{1}} + \frac{1}{U_{2}-V_{2}}\, . \label{eqn:Jdecim4}
\end{align}

A site phase-locking step is performed when $K$ is dominant. The relative phase between the legs is then frozen, $\phi_u=\phi_d+\varphi_K$, so the two clusters of the site merge into a single phase-coherent cluster. We call the resulting site a bowtie site (B). Its on-site Hamiltonian is given by
\begin{align}
    H_B^i = U^B_in_i^2 - \mu_i n_i \label{eqn:HB}
\end{align}
with $U^B_i=\frac{U_i+V_i}{2}$ (note that this definition includes a non-trivial factor of $2$, see a detailed discussion in appendix \ref{app:Bonds}). The inter-site Josephson coupling to the nearest neighboring sites remain the same as before the decimation, similar to the terms in Eq. (\ref{eqn:Hrung}), following the schematic drawings in Fig. \ref{fig:RBRungs}.

In a site charge-locking step, the charge state of a site $i$ with large charging energy is frozen at its ground-state. Note that following Eq. (\ref{eqn:nbar}) we assume $\left|\mu_i\right|\leq \frac{U_i+V_i}{2}$. If $\mu_i$ is close to $0$, the ground state is $n_u=n_d=0$, and the gap to the lowest energy charge excitation is of order $U_i$.
More generally, we define
\begin{align}
    \Delta_i \equiv \left|\frac{1}{2}U_i - \left|\mu_i\right|\right| \label{eqn:chargegap}
\end{align}
to be the local charge gap.
Upon decimation of this site, an effective coupling between its neighboring sites $i\pm 1$ is induced via a second-order hopping. 
The effective Josephson coupling between sites $i-1$ and $i+1$ is given by
\begin{align}
    J_{i-1,i+1}^{eff} \sim \frac{J_{i-1,i}J_{i,i+1}}{\Delta_i} \label{eqn:Udecim}
\end{align}
with $\Delta_i$ the charge gap defined in Eq. (\ref{eqn:chargegap}); the precise expression is derived in appendix \ref{app:S-W} (see Eq. (\ref{appeq:Jeff})). 
This holds for all the connections mentioned in Fig. \ref{fig:RBRungs} -- every two bonds that connect to the same site in rung $i$ create such second-order term.

A distinct scenario occurs at larger values of $\mu_i$ such that the above mentioned gap closes. Specifically, for $\frac{U}{2}<\left|\mu\right|<\frac{U+V}{2}$, the ground state is doubly-degenerate in the subspace $\left|n_u=1,n_d=0\right\rangle\equiv\left|\uparrow\right\rangle$, $\left|n_u=0,n_d=1\right\rangle\equiv\left|\downarrow\right\rangle$. As a result, following an RG step the site does not disappear, but rather change its nature - from R site to what we call a "doublet" (D) site which hosts a spin-like degree of freedom; the on-site Hamiltonian of this D site is a spin Hamiltonian with $K$ serving as a pseudo magnetic field in the $xy$-plane:
\begin{align}
    H_D^i = -\vec{K}_i \cdot \vec{\sigma}^i \label{eqn:HD}
\end{align}
where $\vec{K}_i=K_i\left(\cos\varphi_{K_i},\sin\varphi_{K_i}\right)$
and $\vec{\sigma^i}=\left(\sigma^i_x,\sigma^i_y \right)$
with $\sigma^i_\alpha$ the Pauli matrices.
The inter-site coupling terms are slightly more complicated in the case of D sites and will be discussed from Eq. (\ref{eqn:RDterm}) onward.

To conclude, an R site can undergo four types of steps -- the bond-decimation step, the site phase-locking step, and two variants of the site charge-locking step. The first and the third are the same as in the single chain case described in (ref. to AKPR), while the second and the fourth strongly reflect the  $\mathbb{Z}_2$ symmetry of the ladder.

Moving on to describe B sites, the behavior is very similar to R sites except
there is no site phase-locking step. The bond-decimation step can be described by equations similar to (\ref{eqn:Jdecim1})-(\ref{eqn:Jdecim4}); when R and B sites are merged following a bond-decimation step, as in the case of the $R_b$ unit call shown in Fig. \ref{fig:RBRungs}, the resulting site is a B site with a single superconducting grain.

The site charge-locking step for a B site is simpler - the ground state is always unique\cite{AdditionalSymmetryNote}, so the site is entirely decimated. In the case of an R-B-R chain, when the bowtie is decimated, there are additional second-order Josephson coupling terms $\cos\left(\phi_{u,i-1}-\phi_{d,i+1}\right)$ and $\cos\left(\phi_{d,i-1}-\phi_{u,i+1}\right)$, creating an $R_{r\times}$ site, like the one presented in Fig. \ref{fig:RBRungs}.

Doublet sites pose a slightly more complicated situation. The essential step consists of projecting the Hamiltonian to the low-energy sector of the degenerate ground-state subspace. This yields an effective Hamiltonian that describes their behavior while restricted to this two-states sector.
The on-site Hamiltonian of the doublets, given in Eq. (\ref{eqn:HD}), follows directly from the Josephson coupling on the rungs; their bonds are more complicated, as leading-order terms take them out of the low-energy subspace. The effective inter-site Hamiltonian is therefore dictated by higher order processes and can be derived using Schrieffer-Wolff transformations (see appendix \ref{app:S-W} for a detailed calculation). If the nearest-neighbor site is an R site, the resulting coupling term to a D site is given by 
\begin{align}
   & - \tilde{J}_{i, i+1} \sigma_i^+ e^{i(\phi_{i+1,d}-\phi_{i+1,u})} + h.c. \label{eqn:RDterm} \\
   & \tilde{J}_{i, i+1} \equiv C\frac{J_i^2}{\Omega} \label{eqn:Jtilde}
\end{align}
where $C$ is a factor of order unity that depends on $\frac{V}{U}$ and $\frac{\mu}{U}$ of the original site; for a B site, one has to substitute $\phi_{i+1, d}=\phi_{i+1, u}$. 

Apart from these terms, in an R-D-R configuration there are also next-nearest-neighbors terms which couple two sites at either side of a D site:
\begin{align}
    &-\tilde{J}_{i-1,i+1}^{NNN} \left(\cos(\phi_{i-1,u}-\phi_{i+1,u})+u.d.c.\right) \label{eqn:HRDR} \\
    &-\tilde{J}_{i-1,i+1,\times}^{NNN} \left(\sigma_i^+e^{i(\phi_{i+1,d}-\phi_{i-1, u})}+h.c.+u.d.c.\right) \nonumber
\end{align}
where $u.d.c.$ is the "up-down complement", the operator resulted from taking $u\leftrightarrow d$ and $\sigma^+ \leftrightarrow \sigma^-$; and where
\begin{align}
    \tilde{J}^{NNN}_{i-1,i+1}=C_{=}\frac{J_{i-1}J_i}{\Omega} \nonumber\\
    \tilde{J}^{NNN}_{i-1,i+1,\times}=C_{\times}\frac{J_{i-1}J_i}{\Omega} \label{eqn:JNNN}
\end{align}
where $C_{\times},C_{=}$, similarly to $C$, are factors of order unity that depend on $\frac{V}{U}$ and $\frac{\mu}{U}$.
These terms, as well as the nearest-neighbor terms $\tilde{J}_{i-1,i}$ and $\tilde{J}_{i,i+1}$, are shown in Fig. \ref{fig:DRungs} and their derivation is detailed in appendix \ref{app:S-W}.

As a direct result of Eqs. (\ref{eqn:Jtilde}) and (\ref{eqn:JNNN}), we note that $\tilde{J}_{i-1,i+1}^{NNN},\tilde{J}_{i-1,i+1,\times}^{NNN}$ are both of order $\sqrt{\tilde{J}_{i-1,i} \tilde{J}_{i,i+1}}$. This implies that $\tilde{J}_{i-1, i+1}^{NNN}$ can never be a dominant energy scale: either $\tilde{J}_{i-1,i}$ or $\tilde{J}_{i,i+1}$ will be larger.
The only possible bond-decimating step in $R_d$ and $D_r$ unit cells is, therefore, what happens when $\tilde{J}_{i, i+1}=\Omega$. In such case, $\tilde{J}_{i-1,i+1}^{NNN}\gg\tilde{J}_{i-1,i}$, and the resulting site is an $R_{r\times}$ site -- the process is visualized in the bottom panel of Fig. \ref{fig:DRungs}. This happens since the NNN terms, playing the part of the $J$ and $J_\times$ bonds (blue and cyan in Fig. \ref{fig:RBRungs}), are significantly stronger than the NN terms which control the $R_d$ and $D_r$ unit cells.


When a site adjacent to a D site is decimated, like in the last step depicted in Fig. \ref{fig:JUDProcess}, it is useful to think of $\tilde{J}_{i,i+1}$ as \textbf{second-order} processes. Suppose site $i$ gets site-charge decimated, and $i+1$ is a D site. The new bond energy will be
\begin{align}
    \tilde{J}_{i-1,i+1}^{eff}\propto\frac{\tilde{J}_{i,i+1} J_{i-1, i}^2}{\Omega^2} \label{eqn:UdecimJD}
\end{align}
which matches Eq. (\ref{eqn:Jtilde}), as $\tilde{J}$ scales like $\frac{J^2}{\Omega}$. This behavior will pose some complications that will be addressed in subsection \ref{subsec:Ansatz}.

A site charge-locking step is not possible for a D site, as the system is perfectly symmetric (this is guaranteed by the $\mathbb{Z}_2$ symmetry of the ladder). On the other hand, a site phase-locking step is possible -- if the Hamiltonian of Eq. (\ref{eqn:HD}) has a very large $K$, the doublet will be frozen in a singlet-like state, with a fixed phase-difference between the legs.
In this case the dominant interaction between the remaining sites is, once again, the next-nearest-neighbor term that maintains the ladder structure.

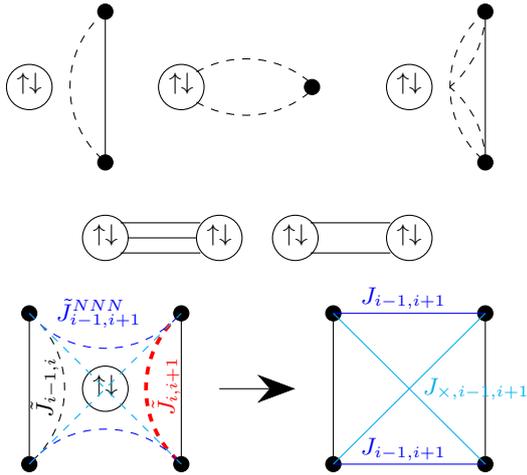
\begin{figure}
\begin{tikzpicture}

\rungD{0}{-1}{0.3}
\draw[dashed] (1, 0) arc (130:230:1.3);
\rungR{1}{0}{-2}{0.1}

\rungD{2}{-1}{0.3}
\draw[dashed] (2.2, -0.8) arc (120:45:1.3);
\draw[dashed] (2.2, -1.2) arc (240:315:1.3);
\rungB{3.72}{-1}{0.1}

\rungD{5}{-1}{0.3}
\draw[dashed] (6, 0) arc (0:-50:1.3);
\draw[dashed] (6, -2) arc (0:50:1.3);
\draw[dashed] (6, 0) arc (130:230:1.3);
\rungR{6}{0}{-2}{0.1}

\rungD{1}{-3}{0.3}
\draw (1.2, -2.8) to (2.25, -2.8);
\draw (1.3, -3) to (2.2, -3);
\draw (1.2, -3.2) to (2.25, -3.2);
\rungD{2.5}{-3}{0.3}

\rungD{3.5}{-3}{0.3}
\draw (3.7, -2.8) to (4.75, -2.8);
\draw (3.7, -3.2) to (4.75, -3.2);
\rungD{5}{-3}{0.3}

\rungR{0}{-6}{-4}{0.1}
\draw[dashed] (0,-4) arc (50:-50:1.3) node[above=0.5cm, right=0.2cm, rotate=90] {$\tilde{J}_{i-1, i}$};
\rungD{1}{-5}{0.3}
\draw[dashed, red, line width=0.5mm] (2,-4) arc (130:230:1.3) node[above=0.5cm, right=-0.2cm, rotate=90] {$\tilde{J}_{i, i+1}$};
\rungR{2}{-6}{-4}{0.1}
\draw[dashed, blue] (0, -4) arc (-140:-40:1.3) node[left=0.4cm] {$\tilde{J}^{NNN}_{i-1,i+1}$};
\draw[dashed, blue] (0, -6) arc (140:40:1.3);
\draw[dashed, cyan] (0, -4) to (2, -6);
\draw[dashed, cyan] (0, -6) to (2, -4);

\draw [-{Stealth[length=5mm]}] (2.5,-5) -- (3.5,-5);

\rungR{4}{-4}{-6}{0.1}
\rungR{6}{-4}{-6}{0.1}
\draw[blue] (4, -4) to (6, -4) node[above=0.2cm, left=0.4cm] {$J_{i-1, i+1}$};
\draw[blue] (4, -6) to (6, -6) node[above=0.2cm, left=0.4cm] {$J_{i-1, i+1}$};
\draw[cyan] (4, -4) to (6, -6) node[above=1.0cm, left=-0.7cm] {$J_{\times,i-1, i+1}$};
\draw[cyan] (4, -6) to (6, -4);

\end{tikzpicture}
\caption{\label{fig:DRungs}
Unit cells involving a doublet (D) site. Top row (left to right): $D_r$, $D_b$, $D_{r\times}$ where the dashed arcs denote $\tilde{J}$-type bonds. Middle row (left to right): $D_d$, $D_{d\times}$; the former (with three lines) generated from $R_{r=}$ sites and the latter from $R_{r\times}$. Bottom row: a bond decimation of an $R_d-D_r$ configuration with D on site $i$. The red bond is decimated, and the resulting $R_{r\times}$ site has interactions dominated by the next-nearest-neighbor blue and cyan terms, not by the black nearest-neighbor term. The resulting $J_{i-1, i+1}$ is equal to $\tilde{J}^{NNN}_{i-1, i+1}$ up to a non-universal factor of order unity, and similarly for $\tilde{J}^{NNN}_{i-1, i+1, \times}$ and $J_{i-1, i+1,\times}$.}
\end{figure}

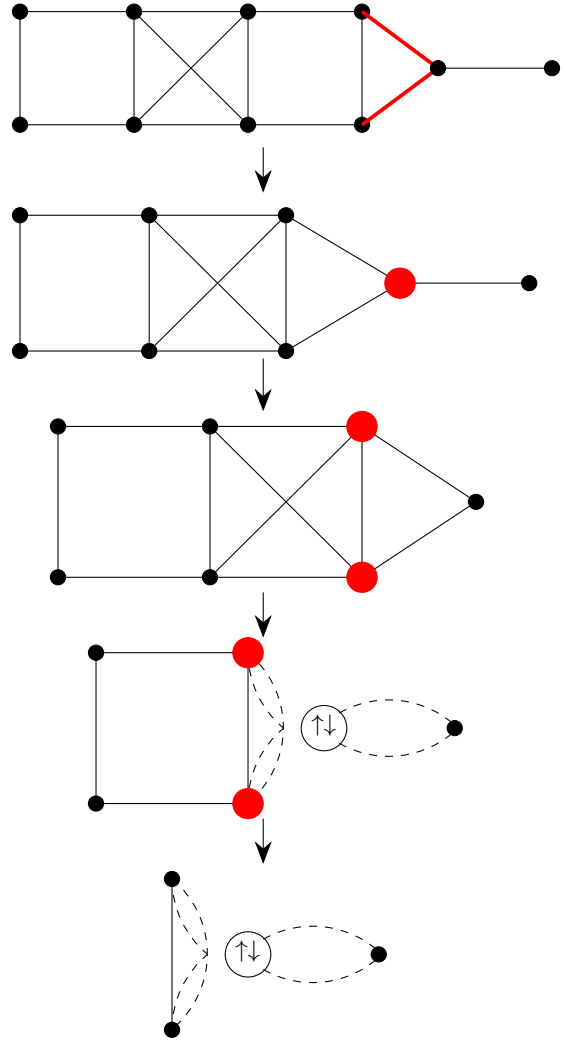
\begin{figure}
\begin{tikzpicture}
\rungR{-0.5}{0}{-1.5}{0.1}
\draw (-0.5, 0) to (1, 0);
\draw (-0.5, -1.5) to (1, -1.5);
\rungR{1}{0}{-1.5}{0.1}
\draw (1, 0) to (2.5, 0);
\draw (1, -1.5) to (2.5, -1.5);
\draw (1, 0) to (2.5, -1.5);
\draw (1, -1.5) to (2.5, 0);
\rungR{2.5}{0}{-1.5}{0.1}
\draw (2.5, 0) to (4, 0);
\draw (2.5, -1.5) to (4, -1.5);
\rungR{4}{0}{-1.5}{0.1}
\draw[red, line width=0.5mm] (4, 0) to (5, -0.75) to (4, -1.5);
\rungB{5}{-0.75}{0.1}
\draw (5, -0.75) to (6.5, -0.75);
\rungB{6.5}{-0.75}{0.1}

\draw [-{Stealth[length=3mm]}] (2.7,-1.8) -- (2.7,-2.4);

\rungR{-0.5}{-2.7}{-4.5}{0.1}
\draw (-0.5, -2.7) to (1.2, -2.7);
\draw (-0.5, -4.5) to (1.2, -4.5);
\rungR{1.2}{-2.7}{-4.5}{0.1}
\draw (1.2, -2.7) to (3, -2.7);
\draw (1.2, -4.5) to (3, -4.5);
\draw (1.2, -2.7) to (3, -4.5);
\draw (1.2, -4.5) to (3, -2.7);
\rungR{3}{-2.7}{-4.5}{0.1}
\draw (3, -2.7) to (4.5, -3.6) to (3, -4.5);
\draw (4.5, -3.6) to (6.2, -3.6);
\filldraw[red] (4.5, -3.6) circle (0.2);
\rungB{6.2}{-3.6}{0.1}

\draw [-{Stealth[length=3mm]}] (2.7,-4.6) -- (2.7,-5.3);

\rungR{0}{-5.5}{-7.5}{0.1}
\draw (0, -5.5) to (2, -5.5);
\draw (0, -7.5) to (2, -7.5);
\rungR{2}{-5.5}{-7.5}{0.1}
\draw (2, -5.5) to (4, -5.5);
\draw (2, -7.5) to (4, -7.5);
\draw (2, -5.5) to (4, -7.5);
\draw (2, -7.5) to (4, -5.5);
\draw (4, -5.5) to (4, -7.5);
\draw (4, -5.5) to (5.5, -6.5) to (4, -7.5);
\filldraw[red] (4, -5.5) circle (0.2);
\filldraw[red] (4, -7.5) circle (0.2);
\rungB{5.5}{-6.5}{0.1}

\draw [-{Stealth[length=3mm]}] (2.7,-7.7) -- (2.7,-8.3);

\rungR{0.5}{-8.5}{-10.5}{0.1}
\draw (0.5, -8.5) to (2.5, -8.5);
\draw (0.5, -10.5) to (2.5, -10.5);
\draw (2.5, -8.5) to (2.5, -10.5);
\draw[dashed] (2.5, -8.5) arc (50:-50:1.3);
\draw[dashed] (2.5, -8.5) arc (180:230:1.3);
\draw[dashed] (2.5, -10.5) arc (180:130:1.3);
\filldraw[red] (2.5, -8.5) circle (0.2);
\filldraw[red] (2.5, -10.5) circle (0.2);
\rungD{3.5}{-9.5}{0.3}
\draw[dashed] (3.7, -9.3) arc (120:45:1.3);
\draw[dashed] (3.7, -9.7) arc (240:315:1.3);
\rungB{5.22}{-9.5}{0.1}

\draw [-{Stealth[length=3mm]}] (2.7,-10.7) -- (2.7,-11.3);

\rungR{1.5}{-11.5}{-13.5}{0.1}
\draw[dashed] (1.5, -11.5) arc (50:-50:1.3);
\draw[dashed] (1.5, -11.5) arc (180:230:1.3);
\draw[dashed] (1.5, -13.5) arc (180:130:1.3);
\rungD{2.5}{-12.5}{0.3}
\draw[dashed] (2.7, -12.3) arc (120:45:1.3);
\draw[dashed] (2.7, -12.7) arc (240:315:1.3);
\rungB{4.22}{-12.5}{0.1}

\end{tikzpicture}
\caption{\label{fig:JUDProcess}
Four RG steps performed one by one on a small ladder. The large energy scale is bold in red. The steps, by order: a bond-decimation step on an $R_b$ site; a site charge-locking step on a $B_b$ site; a site charge-locking step on an $R_b$ site, which results with a    D site; a site charge-locking step on an $R_d$ site, which results with full decimation.}
\end{figure}

The only remaining case we have to discuss is the $D_d$ unit cell - the Hamiltonian coupling two adjacent doublet sites. The interaction between them is derived from the coupling Eq. (\ref{eqn:RDterm}) when site $i+1$ is also projected to the low-energy subspace:

\begin{align}
    -\tilde{J}_i \sigma_i^+ \sigma_{i+1}^- + h.c. \label{eqn:HDD}\, .
\end{align}
This term corresponds to an exchange coupling in an XX-chain. If the original site was $R_{d\times}$, a random anisotropy is also introduced; we will shortly discuss the behavior of the case of such a spin chain. Further discussion regarding similar strongly-disordered spin chains can be found in Ref. \onlinecite{PhysRevB.50.3799}. Some subtleties related to the terms involving D sites are further discussed in appendix \ref{app:RDDR}.

To conclude, we have three types of steps for "regular" ladder sites -- bond-decimation step, that unites two rungs into one; site phase-locking step, that unites the superconducting grains from both legs into one and creates a B-site; and site charge-locking step, that either decimates out the site completely or turns it into a D-site. For B sites there is no site phase-locking step, and for D sites there is no site charge-locking step. The possible unit cells involving R and B sites can be seen in Fig. \ref{fig:RBRungs}, and unit cells involving D sites can be seen in Fig. \ref{fig:DRungs}. An example for a short sequence of steps is shown in Fig. \ref{fig:JUDProcess}.

Every possible step has a relevant energy scale - either charging (for site charge-locking) or Josephson (for site phase-locking and for bond-decimation). $\Omega$ is defined as the maximal energy scale across the whole ladder, and we repeatedly take the relevant Hamiltonian term, assume that the related degree of freedom is frozen, and write the effective Hamiltonian.

This process can be numerically simulated; alternatively, the distribution of the parameters can be analyzed and one can write master equations to describe their flow. In the following sections we will use both methods to analyze the strongly-disordered Josephson ladder and describe its phase diagram.

\subsection{Distributions and Master Equation}
\label{subsec:MasterEqns}
In many perturbative RG methods, the disorder is parametrized by a single parameter, usually the variance of some value. However, it is the nature of strong disorder and broad distributions not to be well-characterized by a single parameter, as a Gaussian approximation might not be valid. To study the flow under strong disorder we have to study the flow of the entire distribution.

In this subsection we will mostly discuss the behavior of $R$-sites. The further assumptions we will require are explained in detail in the next subsection. All the assumptions mentioned below were numerically justified unless explicitly stated otherwise.

One assumption that will already be necessary here is the assumption that $\frac{V}{U}$ is approximately a constant that we will denote $v$. This is justified because when two adjacent sites merge into one in a bond-decimation, $\frac{V}{U}$ is given by a weighted average dictated by Eq. (\ref{appeq:veff}); assuming, without loss of generality, $\frac{V_1}{U_1}<\frac{V_2}{U_2}$, this yields
\begin{align}
\frac{V_1}{U_1} < \frac{V_{eff}}{U_{eff}} < \frac{V_2}{U_2}. \label{eqn:vAssumption}
\end{align}
Hence, a disorder in $\frac{V}{U}$ will be irrelevant (in the RG sense); see appendix \ref{app:Bonds} for a more detailed discussion and a derivation of this result. Under this assumption, Eqs. (\ref{eqn:Jdecim3}) and (\ref{eqn:Jdecim4}) reduce to
\begin{align}
U_{eff}^{-1} \approx U_1^{-1} + U_2^{-1}\, . \label{eqn:Ueffapprox}
\end{align}

Eq. (\ref{eqn:Ueffapprox}) suggests that a useful dimensionless parametrization for $U$ would be 
\begin{align}
\zeta_i=\frac{2\Omega}{U_i}    
\end{align}
where the factor $2$ in the numerator is for simplicity of the following algebra.
Similarly, Eq. (\ref{eqn:Udecim}) leads to introducing
\begin{align}
    \beta_i=\log\left(\frac{\Omega}{J_i}\right). \label{eqn:betaJ}
\end{align}
With these parametrizations and the assumption mentioned above, we have three simple equations describing the effective parameters following bond decimations:
\begin{align}
\beta_{eff} =& \beta_1+\beta_2 \label{eqn:betastep}\\
\zeta_{eff} =& \zeta_1 + \zeta_2 \label{eqn:zetastep}\\
\bar{n}_{eff} =& \bar{n}_1 + \bar{n}_2\,. \label{eqn:nstep}
\end{align}
We shall further assume that $\beta_i$ and $\zeta_i$ are independent. However, we can not assume that $\zeta_i$ and $\bar{n}_i$ are independent, as we require $\Delta_i<\Omega$. Following Eq. (\ref{eqn:chargegap}) this requirement couples the two. Rewriting Eq. (\ref{eqn:chargegap}) in terms of $\zeta$ and $\bar{n}$, where $\bar{n}$ was defined in Eq. (\ref{eqn:nbar}), it implies that $\zeta_i$ is bounded by a critical value $\zeta_c\left(\left|\bar{n}_i\right|\right)$:
\begin{align}
\zeta_i > \zeta_c\left(n_i\right) \equiv\left|1-2(1+v)\left|\bar{n}_i\right|\right| \label{eqn:zeta_n}
\end{align}

We therefore introduce two variables of the RG flow which are distribution functions: $f(\zeta, \bar{n})$ and $g(\beta)$. The domain of these functions is $\zeta\in\left[0,\infty\right),\bar{n}\in\left[-\frac{1}{2},\frac{1}{2}\right]$, further restricted to the points that satisfy Eq. (\ref{eqn:zeta_n}); $g$ is defined for $\beta\in\left[0,\infty\right)$.
In what follows, we study
the behavior of these distributions as a function of our flow parameter, $\Gamma$, defined by 
\begin{align}
    \Omega = \Omega_0 e^{-\Gamma} \label{eqn:Gamma},
\end{align}
so we will often write $f(\zeta,\bar{n},\Gamma)$ and $g(\beta, \Gamma)$.

There is another distribution we will have to describe, for the parameters $K$ and $\varphi_K$. While they have independent distributions, and effectively $\varphi_K$ is widely distributed in $\left[0,2\pi\right]$, it makes sense to use the parametrization $\kappa=\frac{K e^{i\varphi_K}}{\Omega}$ as Eq. (\ref{eqn:Jdecim1}) implies that
\begin{align}
\kappa_{eff} = \kappa_1 + \kappa_2\,. \label{eqn:kappastep}
\end{align}
The parameter $\kappa$ is a complex number in the unit circle;   we denote its distribution by $h(\kappa)$.

The four equations (\ref{eqn:betastep}), (\ref{eqn:zetastep}), (\ref{eqn:nstep}) and (\ref{eqn:kappastep}) dictate the behavior of the different RG steps. To complete the picture, we introduce the rates of these steps. If we let $\Omega\to\Omega e^{-d\Gamma}$, all three parameters ($\zeta$, $\beta$, $\kappa$) get rescaled and some of them are pushed through the energy cutoff $E=\Omega$. We define $\gamma_f, \gamma_g$ and $\gamma_h$ to be the rates for site charge-locking, bond-decimation and site phase-locking steps (correspondingly). For $\gamma_g$ and $\gamma_h$, these have fixed values
\begin{align}
    \gamma_g \equiv& g(0) \\
    \gamma_h \equiv& 2\pi\cdot h(1). \label{eqn:kappa0}
\end{align}
Both of these are flow rates: while changing $\Gamma\to \Gamma+d\Gamma$, there are $\gamma_g d\Gamma$ bond-decimation steps and $\gamma_hd\Gamma$ site phase-locking steps. The equivalent expression for $f$ is more complicated: 
\begin{align}
    \gamma_f \equiv \int_{-\frac{1}{2}}^{\frac{1}{2}} \zeta_c\left(n\right) f(\zeta_c\left(n\right), n) dn \label{eqn:gammafdef}
\end{align}
which arises from the restriction in Eq. (\ref{eqn:zeta_n}).
We further define $\gamma_d$ such that $\gamma_f\gamma_d$ of the steps give  rise to doublets and $\gamma_f(1-\gamma_d)$ totally decimate the site.

Using these definitions, it is relatively straightforward to derive a master equation for $g(\beta, \Gamma)$. The rate of its flow is dictated by steps that give rise to Eq. (\ref{eqn:betastep}) - namely, turning two bonds into a single one following the decimation of an intermediate site. The obvious contribution to this is $\gamma_f(1-\gamma_d)$, but there are additional processes that lead to the same result. For example, after an inter-site bond decimation uniting two different SC islands, two R sites with $\Delta < \Omega$ (with $\Delta$ the charge gap defined in Eq. (\ref{eqn:chargegap})) might create a single site with $\Delta>\Omega$. Such a site will be immediately charge-decimated, therefore the rate of this process also contributes to said rate -- we encapsulate all these processes to a single rate parameter which we denote $\tilde{\gamma}_f$.

We similarly denote the rate of bond decimations that remove bonds with $\beta\approx 0$ by $\tilde{\gamma}_g$. For the original ladder, $\tilde{\gamma}_g=\gamma_g$, but when D sites are introduced, the situation is more complicated -- as we further discuss in the next subsection. 
Using these two rates, one can write a master equation for $g$:


\begin{align}
    \frac{\partial g}{\partial\Gamma}=\frac{\tilde{\gamma}_g}{\gamma_g}\frac{\partial g}{\partial\beta} + \tilde{\gamma}_f(g * g) + (\tilde{\gamma}_g-\tilde{\gamma}_f)g \label{eqn:gmaster}
\end{align}
where
\begin{align}
    (g * g)(\beta) \equiv \int_0^\infty g(\beta^\prime) g(\beta-\beta^\prime)d\beta^\prime
\end{align}
is the auto-convolution of $g$.

The flow equations for $f$ and $h$ will require more terms accounting for the cases of $\zeta, \bar{n}$ that do not follow Eq. (\ref{eqn:zeta_n}) or $\kappa$ with $\left|\kappa\right|>1$. But the main idea is the same -- we take into account the new sites created, the old sites decimated, the re-definition of $\beta, \zeta$ due to the change in $\Omega$, and the re-scaling of the distributions due to the change in the total number of sites.

Such equations can be solved numerically. We have chosen not to do so, and instead focus on finding an ansatz that allows us to turn them into simpler equations for specific parameters, along with numerical simulations of the single-step logic described in subsection \ref{subsec:Elementary} to test the applicability of said ansatz.


\subsection{Ansatz}
\label{subsec:Ansatz}

As noted above, constructing the explicit master equations for the distributions $f$ and $h$ is more complicated, since simple bond decimations and site phase-locking steps might generate new sites with $\Delta_i>\Omega$ leading to follow-up decimations and more complicated processes. The details of these processes can be extracted from $f(\zeta, \bar{n})$ but they depend on the exact unit cell (say, $R_{r\times}$ vs. $R_{r=}$ vs. $R_b$); putting them into a single equation is neither solvable nor illuminating.
Instead, we pose an Ansatz on the distribution functions. This ansatz will \textbf{not} generally hold; however, it is a useful baseline to discuss even in cases where it does not hold.

We start by reducing the number of distinct distribution functions required to describe the asymptotic RG flow. For example, the distribution of $\kappa$ which we have defined as a single function $h(\kappa)$ might in principle be different for R sites and D sites. However, as indicated by numerical simulations (see Fig. \ref{fig:kappa} below),
$h$ flows to a single semi-stable distribution; we hence assume that the distributions of $h$ for $R$ and $D$ are the same. \bcol{[Perhaps add:] Numerical evidence to this claim, and other claims in this subsection, can be found in appendix \ref{app:Ansatz}.}

We further assume that all the bonds between R and B sites come from the same distribution $g(\beta)$.
To account for bonds that include D sites, we note that 
following Eq. (\ref{eqn:UdecimJD}) one can assume the distribution of $\tilde{J}$ to be such that 
\begin{align}
\tilde{J}=\Omega e^{-2\beta} \label{eqn:betaJtilde}
\end{align}
with $\beta$ drawn from the same distribution $g(\beta)$. This definition means that Eq. (\ref{eqn:betastep}) still holds for a site decimation, even when it includes $\tilde{J}$. It also means that a step like the one described in the bottom row of Fig. \ref{fig:DRungs} does not change $\beta$: two $\tilde{J}$ bonds, one with $\beta=0$ and one with $\beta=\beta_1>0$, turn into a single $J$ bond with $\beta=\beta_1$.
This assumption is necessarily an approximation. If at some point we have two bonds, $J$ and $\tilde{J}$, which correspond to the same $\beta$ -- namely, we have $J=\Omega e^{-\beta}$ and $\tilde{J}=\Omega e^{-2\beta}$ -- after the flow continues and we get to $\Omega^\prime = \Omega e^{-c}$, we use Eqs. (\ref{eqn:betaJ}) and (\ref{eqn:betaJtilde}) to get

\begin{align}
\beta_J&=\log\left(\frac{\Omega^\prime}{J}\right)=\beta+c \\
\beta_{\tilde{J}}&=\frac{1}{2}\log\left(\frac{\Omega^\prime}{\tilde{J}}\right)=\beta+\frac{c}{2} \,.\label{eqn:betaJtildeDisc}
\end{align}
This behavior implies that the distributions of $\beta_J$ and $\beta_{\tilde{J}}$ can not be evolving in the exact same way. However, it turns out that the rate at which $J$ bonds turn into $\tilde{J}$ bonds and vice versa does not allow these distributions to get very far apart, and we shall assume that there is a single $g(\beta)$ distribution for all bond types.

We next turn to describe the distributions of the site-parameters $U, V, \mu$ for R sites, and $U^B, \mu$ for B sites (as presented in Eq. (\ref{eqn:HB})). As in the previous section, we assume that $\frac{V}{U}=v$ is a constant. The distributions of $U, U^B$ and $\mu$ cannot be the same, as the equation for the gap (\ref{eqn:chargegap}) and the implied restriction (\ref{eqn:zeta_n}) are only valid for an R site. We will assume that these distributions share parameters, in a way that will be explained shortly.

Following the above discussion, we assume that all $\beta$s and $\kappa$s are drawn from the same distribution -- but what is this distribution? For $\beta$ the answer is simple -- we assume it to be an exponential, characterized by a single variable $g_0$:

\begin{equation}
g(\beta) = g_0 e^{-g_0\beta} \, .\label{eqn:gbeta}
\end{equation}
This distribution maintains its structure under steps of the form of Eq. (\ref{eqn:betastep}), as can be seen by substituting in Eq. (\ref{eqn:gmaster}).

The behavior of $h(\kappa)$ is slightly more complicated. $\kappa$ undergoes two processes -- one is addition according to Eq. (\ref{eqn:kappastep}), and the other is rescaling $\kappa \to \kappa e^{d\Gamma}$ due to the change of cutoff. However, it will turn out that we are mostly interested in $\gamma_h$, which is directly related to $h(1)$ by Eq. (\ref{eqn:kappa0}).

The first process, addition, has a fixed point with $h(0)\approx 0.41$ and $h(1) \approx 0.22$. An illustration and the numerical result can be found in Fig. \ref{fig:kappa}. The second process is rescaling as a result of restoring the cutoff (see the gray area in the figure). After normalization of this region we get lower values of $h(0)-h(1)$ and the distribution is closer to a uniform one.

The balance between these two processes is controlled by $\gamma_g$, which determines the rate of bond decimations (causing the addition process) compared to the rescaling rate. As we will see later, the most important question is whether $\gamma_h$ is larger or smaller than $1$, and we find that $\gamma_h>1$ in all cases. For $\gamma_g\to\infty$ there is no rescaling and $\gamma_h=2\pi h(1)\approx 1.38$, and for $\gamma_g\to0$ there is no addition and $h(1)=\frac{1}{\pi}$ and $\gamma_h=2$. Both boundary cases are shown in Fig. \ref{fig:kappa}b.

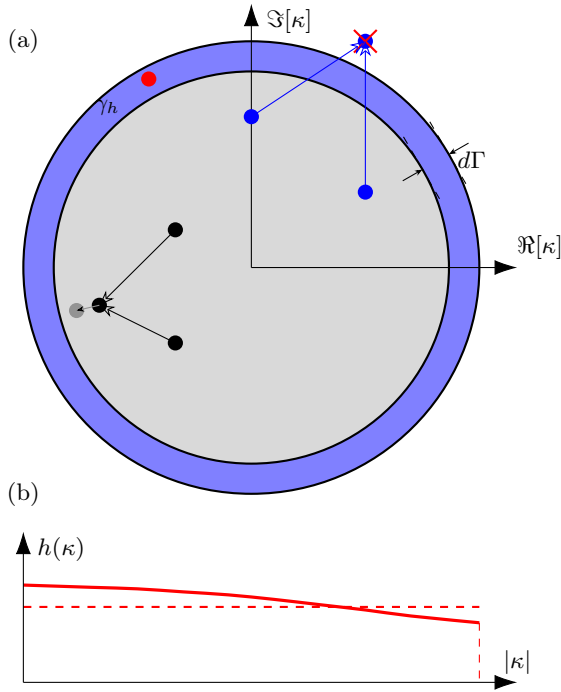
\begin{figure}
\begin{tikzpicture}
\node[] at (-3, 3) {(a)};

\fill[blue!50] (0,0) circle (3.0cm);
\fill[gray!30] (0,0) circle (2.6cm);
\draw [thick] (0, 0) circle (3.0) node[right=2.9cm, above=1.2cm] {$d\Gamma$};
\draw [thick] (0, 0) circle (2.6) node[left=1.9cm, above=1.9cm] {$\gamma_h$};
\draw [-{Stealth[length=1mm]}] (2.3*0.866,2.3*0.5) -- (2.6*0.866,2.6*0.5);
\draw [-{Stealth[length=1mm]}] (3.3*0.866,3.3*0.5) -- (3.0*0.866,3.0*0.5);
\draw[dashed] (2.6*0.866 - 0.5*0.5, 2.6*0.5 + 0.5*0.866) to (2.6*0.866 + 0.5*0.5, 2.6*0.5 - 0.5*0.866);
\draw[dashed] (3.0*0.866 - 0.5*0.5, 3.0*0.5 + 0.5*0.866) to (3.0*0.866 + 0.5*0.5, 3.0*0.5 - 0.5*0.866);

\draw [arrows = {-Latex[width=2mm, length=3mm]}] (0,0) -- (3.5,0) node[right=0.3cm, above=0cm] {$\Re[\kappa]$};
\draw [arrows = {-Latex[width=2mm, length=3mm]}] (0,0) -- (0,3.5) node[right=0.5cm, above=-0.5cm] {$\Im[\kappa]$};

\fill (-1, 0.5) circle (0.1);
\fill (-1, -1) circle (0.1);
\fill (-2, -0.5) circle (0.1);
\fill[gray!85] (-2.3, -0.57) circle (0.1);
\draw[black] [-{Stealth[length=2mm,open]}] (-1,0.5) to (-2,-0.5);
\draw[black] [-{Stealth[length=2mm,open]}] (-1, -1) to (-2, -0.5);
\draw[gray!85] [-{Stealth[length=1mm, open, black]}] (-2, -0.5) to (-2.3, -0.57);

\fill[blue] (0, 2) circle (0.1);
\fill[blue] (1.5, 1) circle (0.1);
\fill[blue] (1.5, 3) circle (0.1);
\draw[red, thick] (1.35, 2.85) to (1.65, 3.15);
\draw[red, thick] (1.65, 2.85) to (1.35, 3.15);
\draw[blue] [-{Stealth[length=2mm,open]}] (0, 2) to (1.5, 3);
\draw[blue] [-{Stealth[length=2mm,open]}] (1.5, 1) to (1.5, 3);

\fill[red] (-1.35, 2.5) circle (0.1);

\node[] at (-3, -3) {(b)};

\draw[very thick, red] plot[smooth] coordinates {(-3+6*0, 1.29-5.5) (-3+6*0.15, 1.26-5.5) (-3+6*0.25,1.24-5.5) (-3+6*0.35, 1.20-5.5) (-3+6*0.45,1.16-5.5) (-3+6*0.55,1.10-5.5) (-3+6*0.65,1.03-5.5) (-3+6*0.75,0.96-5.5) (-3+6*0.85,0.88-5.5) (-3+6*1.0,0.79-5.5)};
\draw[red, dashed] (3, -4.71) to (3, -5.5);
\draw[thick, dashed, red] (-3, -4.5) to (3, -4.5);
\draw [arrows = {-Latex[width=2mm, length=3mm]}] (-3,-5.5) -- (3.5,-5.5) node [right=0.0cm, above=0.0cm] {$\left|\kappa\right|$};
\draw [arrows = {-Latex[width=2mm, length=3mm]}] (-3,-5.5) -- (-3,-3.5) node[right=0.5cm, above=-0.5cm] {$h(\kappa)$};

\end{tikzpicture}
\caption{
An illustration of the distribution of $\kappa$.\\
(a) The blue area marked with $\gamma_h$ is the region $\Omega e^{-d\Gamma}<K<\Omega$, or $1-d\Gamma<\left|\kappa\right|<1$, soon to undergo a site phase-locking step as the flow goes on. Two pairs of points, in black and in blue, show the addition process. The blue pair will be immediately site phase-locking decimated, as they landed outside of the circle; the black remains in the circle and will be rescaled to the adjacent gray point after the blue ring will be decimated. Another point, drawn in red, will be site phase decimated soon, as it is in the blue area. \\
(b) A numerical approximation of the stable distribution $h(\kappa)$ in the case of almost no rescaling at all ($\gamma_g\gg 1$); the dashed reference line is the uniform case $h(\kappa)=\frac{1}{\pi}$ ($\gamma_g \ll 1$).
\label{fig:kappa}
}
\end{figure}

The Ansatz for $f\left(\zeta,\bar{n}\right)$ is more complicated, but it is just as direct. Due to Eqs. (\ref{eqn:zetastep}) and (\ref{eqn:nstep}), following a previous work\cite{PhysRevB.81.174528} we assume that $\bar{n}$ has a uniform distribution and $\zeta$ has an exponential distribution, but the values must obey Eq. (\ref{eqn:zeta_n}). Using the step function $\chi(x)$ we write
\begin{align}
f(\zeta, \bar{n})=f_c e^{-f_0 \zeta} \chi(\zeta-\left|1-2(1+v)\left|\bar{n}\right|\right|) \label{eqn:fzetan}
\end{align}
where $f_c=\frac{f_0^2(1+v)}{2-e^{-f_0v}-e^{-f_0}}$ is the normalization constant.
For B sites the situation is similar, only we substitute $v=0$ in (\ref{eqn:zeta_n}) to get
\begin{align}
f(\zeta, \bar{n})=\frac{f_0^2}{1-e^{-f_0}} e^{-f_0\zeta} \chi\left(\zeta-(1-2\left|\bar{n}\right|)\right). \label{eqn:fzetanb}
\end{align}

This concludes the assumptions required for our Ansatz. We parameterize $f$ and $g$ with $f_0$ and $g_0$ and investigate their flow, while $h$ is parameterized by an almost constant $\gamma_h$. These three parameters tell us the overall parameter distribution of each site type. The only remaining variables required to fully characterize the system and its flow equations is the probability for a site to be of a given type. We define the type fractions $r, b, s$ for regular, bowtie and doublet (spin-like) sites, correspondingly; they obey $r+b+s=1$. We assume that the types of adjacent sites are not correlated.

Under the assumptions of the Ansatz, we can calculate additional quantities. The most important ones are the process rates. When we take $\Omega\to\Omega e^{-d\Gamma}$, we know that $\gamma_hd\Gamma$ of the sites undergo a site phase-locking step. As to bond-decimation steps, the rate for $J$ terms is $\gamma_gd\Gamma$, but for $\tilde{J}$ terms the rate is $\frac{1}{2} \gamma_g d\Gamma$. The overall rate is
\begin{align}
    \tilde{\gamma}_g = (1-s)^2 \gamma_g + \frac{1-(1-s)^2}{2} \gamma_g = \frac{1+(1-s)^2}{2}\gamma_g
\end{align}
with $\tilde{\gamma}_g$ matching the one of Eq. (\ref{eqn:gmaster}).

For site charge-locking steps the rate can be calculated using Eq. (\ref{eqn:gammafdef}), yielding $\gamma_f, \gamma_d$:
\begin{align}
    \gamma_f \gamma_d=&\frac{1-e^{-f_0 v}(1+f_0v)}{2-e^{-f_0}-e^{-f_0 v}} \\
    \gamma_f(1-\gamma_d)=&\frac{1-e^{-f_0}(1+f_0)}{2-e^{-f_0}-e^{-f_0 v}}\,.
\end{align}
Most notably, for $f_0\to \infty$ the rate does not diverge; instead, $\gamma_f \to 1$, so even for diverging $f_0$ the rate of site charge-locking steps is constant. We can similarly define $\gamma_f^b$ for the bowtie sites -- the resulting rate is
\begin{align}
    \gamma_f^b = \frac{1-e^{-f_0}(1+f_0)}{1-e^{-f_0}} \label{eqn:gammafb}
\end{align}
for which we also have $\gamma_f^b \to 1$ for $f_0 \to \infty$.

The parameters $\gamma_f^b, \gamma_f, \gamma_d$ and $\gamma_h$ can be numerically calculated for any given distribution. Therefore, even if the ansatz does not hold, they still carry their original meanings, and we can follow their flow and observe their numerical behavior. When writing the flow equations, we use these parameters and not substitute the ansatz-derived quantities, so the results are not dependent on the validity of the ansatz.

\subsection{Details of the Numerical Simulations}
\label{subsec:SimulationDetails}
As mentioned above, we would like to accompany the analytical analysis with numerical results. The numerical simulations start with a randomly sampled ladder, which assumes the basic structure of Eqs. (\ref{eqn:Hsum})-(\ref{eqn:Hrung}), made out of R sites. We first randomly select the initial values of $U, V, \mu, J, K$ and $\varphi_K$. Then, the RG steps described in subsection \ref{subsec:Elementary} are performed iteratively on the locally dominant energy scales. In this section we will shortly discuss the specifics of the ladder sampling.

It is known that the exponential distributions of $\beta$ [Eq. (\ref{eqn:gbeta})], and of $\zeta$, given by the $e^{-f_0\zeta}$ term of Eq. (\ref{eqn:fzetan}), are stable fixed points: systems that start with random $\zeta, \beta$ distributions converge rather quickly to the exponential behavior - see figure 3b in Ref. [\onlinecite{PhysRevLett.93.150402}]. Therefore, we would like to already draw $\zeta$ and $\beta$ from an exponential distribution, to avoid the noise involved in the process of converging to the Ansatz.

For $V$, despite Eq. (\ref{eqn:vAssumption}), we let the ratio $v=\frac{V}{U}$ vary in some given range. It turns out that this range does not qualitatively change the behavior of the RG flow, in agreement with the discussion after Eq. (\ref{eqn:vAssumption}). Generally, the exact value of $v$ does not change the behavior of the system.

For $K$ and $\mu$ the convergence to a stable distribution is not guaranteed; moreover, as we show in subsection \ref{subsec:SC-BGTrans}, there is actually a phase transition in the system that rises from the inaccuracy of the ansatz. Therefore, for these two parameters we choose dimensionless parameterizations, $\bar{n}$ defined in Eq. (\ref{eqn:nbar}) and $\kappa$ defined just before Eq. (\ref{eqn:kappastep}), and draw them from narrow distributions, $\bar{n}\in[-n_0,n_0]$ and $\left|\kappa\right|\in [0,\kappa_0]$, corresponding to a weak-disorder limit.

The distribution of $\varphi_K$ is slightly more complicated. If we assume that $\varphi_K$ arises from a magnetic flux, the distribution of the flux determines $\Delta\varphi_{K,i}\equiv\varphi_{K,i+1}-\varphi_{K,i}$. In the limit of weak disorder, the distribution of $\Delta\varphi_{K,i}$ is narrow, which leads to a finite correlation length $\ell_K\approx \frac{1}{\langle\Delta \varphi_K^2\rangle}$. This length scale necessarily flows down with the RG flow, hence we always let $\ell_K=0$ and totally randomize the phase between adjacent sites.

\subsection{Spin Chain Behavior}
\label{subsec:SpinChain}

We are yet to describe the $D_d$ and $D_{d\times}$ unit cells (shown in Fig. \ref{fig:DRungs}), which control the behavior of the generated spin-chain segments. We shall describe the Hamiltonians but not dive into their analysis.

The scenario leading to a spin chain is when, following several RG steps, two or more adjacent sites undergo a site charge-locking step that leaves them as spins.
After two adjacent R sites undergo such step, the resulting interaction is given by Eq. (\ref{eqn:HDD}). Along with the on-site term from Eq. (\ref{eqn:HD}) we get the effective unit cell Hamiltonian:
\begin{align}
H_{D_d}=-\vec{K} \cdot\vec{\sigma_i} - \tilde{J} (\sigma_i^+\sigma_{i+1}^- + \sigma_i^- \sigma_{i+1}^+) \label{eqn:Hspins}
\end{align}

If the unit cell before the steps was $R_{r\times}$, we have to take into account the phase along the diagonals. A careful calculation shows that the inter-site terms of the $R_{r\times}$ site are
\begin{align}
    H_{R_{r\times}}^J=&-J_i\left[\cos(\phi_{i,u}-\phi_{i+1,u})+u.d.c.\right] \nonumber \\
    &-J_{i\times} \left[\cos(\phi_{i,u}-\phi_{i+1,d}-\varphi_J)+u.d.c.\right]
\end{align}
with $\varphi_J$ denoting a phase-difference between the rungs, and it tends to get random just like $\varphi_K$.
When we project this Hamiltonian to the spin subspace, this results additional terms of the form
\begin{align}
    H_{D_{d}}^J = \tilde{J} \left(e^{i\varphi_J} \sigma_i^+ \sigma_{i+1}^+ + h.c.\right) \\
    H_{D_{d}}^{J_\times} = \tilde{J}_\times \left(e^{i\varphi_J} \sigma_i^+ \sigma_{i+1}^- + h.c.\right)
\end{align}
as well as corrections to $K$. Such terms create an anisotropy in $J$.
In the general case, we get the following Hamiltonian:
\begin{align}
    H_{D_{d\times}}^i = (\tilde{J}_i+\tilde{J}_{i\times}) \sigma_i^x \sigma_{i+1}^x + (\tilde{J}_i-\tilde{J}_{i\times}) \sigma_i^y\sigma_{i+1}^y
\end{align}
where the $x, y$ axes are \textbf{rotated} by $\varphi_J$. If $J=\tilde{J}$, as happens when $R_{r\times}$ is generated after a site charge-locking step of a B site, this yields the Ising Hamiltonian; if $J_\times < J$, as happens after a site-locking step of a D site, this is an XY Hamiltonian.

In addition to these terms, there are higher-order terms of the form $\sigma_i^z \sigma_{i+1}^z$, but in the limit of strong disorder $J\ll \Omega$ their coefficient is very small compared to $\tilde{J}$.

Accounting for all these possibilities, the spin chain is an XY model with random anisotropy and random fields, and it can undergo two types of steps -- site steps (if $K$ is large) and bond-decimation steps (if $J$ is large).
The resulting behavior of such disordered spin chains is discussed in earlier literature\cite{PhysRevB.50.3799, PhysRevB.51.6411}.

In the case where the RG flow does not lead to a spin chain, but rather to a local set of spin sites surrounded by many R and B sites, we do not focus on the spin sites themselves, but on the long-range terms between the surrounding rungs, as depicted in Fig. \ref{fig:DRungs}. 
It turns out that these long-range terms can not be calculated from the short-range terms. For the case of an $R_d-D_d-D-r-R_r$ chain the next-next-nearest-neighbors terms between the R sites can be roughly approximated by $J_{i-1, i+2} \approx \sqrt{\frac{\tilde{J}_{i-1,i}\tilde{J}_{i,i+1}\tilde{J}_{i+1,i+2}}{\Omega}}$. The source of this approximation, along with a detailed discussion regarding the behavior of these long-range terms, is presented in appendix \ref{app:RDDR}; we continue with this approximation and note that our description for the mixed state between spins and rungs is not accurate -- so we shall restrict ourselves to describing phases dominated by one type.



\section{Results}
\label{sec:Results}
After we presented the model and studied the possible steps of the RG flow, we now turn to the results. Below we present results from two different types of calculations: one type is numerical, based on creating a specific realization of the system and running the RG steps; the other type is analytical results regarding the flow of the parameters describing the system from the previous section.

The numerical results presented here will be for $\kappa_0\ll \bar{n}_0$. This selection gives the richest phase diagrams, with three different phases clearly visible, as we discuss below in subsection \ref{subsec:SC-BGTrans}.

\begin{figure}
\includegraphics[width=0.5\textwidth]{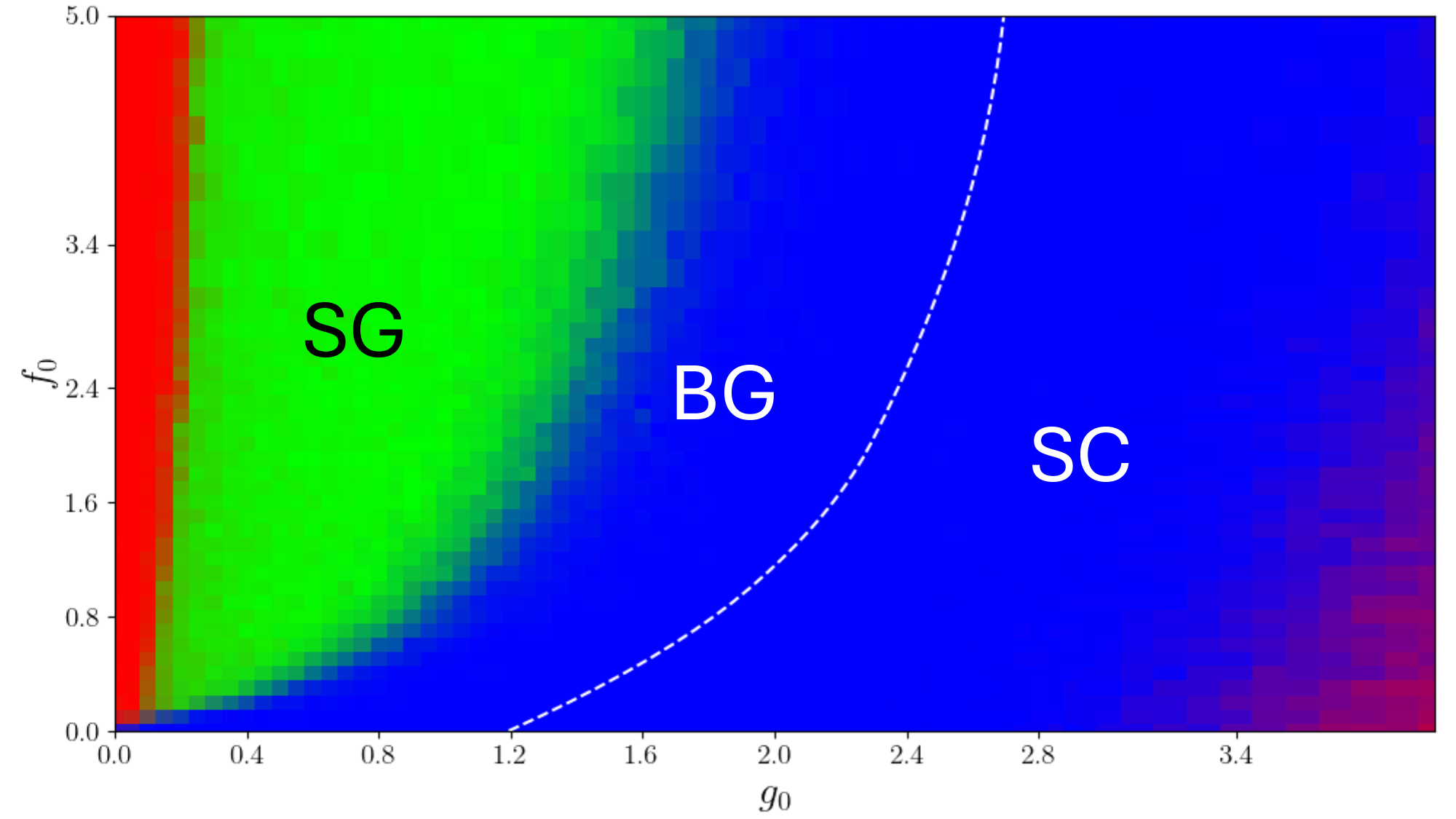}
\caption{
\label{fig:heatmap}
A phase diagram of the system as a function of $f_0$ and $g_0$. Each tile is colored in RGB such that its Red value is $r$, its Green value is $s$ and its Blue value is $b$, corresponding to $r, s, b$ at the end of the flow. The white line is the transition between the Bose glass (BG) and the superconductor (SC), while the boundary between the blue and green regions marks a transition from the BG to a spin glass (SG) insulator. The white line splits the region based on $f_0(\Gamma)$ and $g_0(\Gamma)$ -- it separates the region where $f_0(\Gamma)>g_0(\Gamma)$, indicating $f_0\to \infty$, from the region where $f_0(\Gamma)<g_0(\Gamma)$, indicating $f_0\to 0$.
Using the quantities defined in \ref{subsec:SimulationDetails}, this diagram was generated with $\bar{n}_0=0.11$, $\kappa_0=1.2\cdot 10^{-3}$, $\ell_K=0$ (no correlations in $\varphi_K$) and $v\in[0.5,0.7]$.
}
\end{figure}
\subsection{Phase Diagram}
A phase diagram of the system as a function of $f_0$ and $g_0$ can be seen in Fig. \ref{fig:heatmap}. As mentioned above, we focus on the $\kappa_0 \ll \bar{n}_0$ regime which provides the richest phase diagram.
This figure presents the result of a numerical simulation -- values of $\beta, \zeta, \kappa$ and $\bar{n}$ were drawn from distributions that partially match the Ansatz described in Sec. \ref{subsec:Ansatz} to create a ladder made of R sites like the one described in Eq. (\ref{eqn:Hsum}).

The parameter selection process was supposed to model the case of flow from a weak-disorder parameter point, following the details in \ref{subsec:SimulationDetails}.

The figure shows three phases -- a spin chain, dominated by D sites, and two distinct single-chain phases dominated by B sites. Notably, none of the stable phases maintains the original ladder structure.

The two single-chain phases match the known phases of a single-chain system for the case of a generic disorder\cite{PhysRevLett.100.170402, PhysRevB.81.174528}: the one with $f_0\to 0$ is a superconductor, and the one with $f_0 \to \infty$ is a Bose glass insulator phase. The spin chain phase realizes a distinct type of insulator where the inter-chain degree of freedom is a random spin chain with spins strongly coupled to a random field; i.e., a spin glass phase which is distinct from the spin singlet phases found in similar systems without the ladder structure\cite{PhysRevLett.100.170402}.
The spin-spin interactions prohibit the transfer of charge, hence this phase is an insulator that does not conduct longitudinally. 

The Bose glass is an intermediate phase. While transverse super-current flows through it, the system is insulating to longitudinal currents along the chains\cite{PhysRevB.81.174528}. The superconductor is superconducting in all modes, both longitudinal and transverse, dominated by strong Josephson couplings while the charging energies are irrelevant. However, while the anti-symmetric sector is fully phase-locked exhibiting long-range 
SC order, the symmetric sector exhibits quasi-long-range SC order.

\subsection{Analysis of the RG flow equations}
We now turn to understand the RG flow characterizing the system, and its possible stable states. Fig. \ref{fig:heatmap} shows the results of a numerical simulation, following the guidelines of subsection \ref{subsec:SimulationDetails}. We stress that this choice of initial distributions does not fully match the Ansatz  -- specifically, regarding $n$ and $K$, following the above discussion.

We start by writing the flow equation for $g_0$.
Substituting Eq. (\ref{eqn:gbeta}) in Eq. (\ref{eqn:gmaster}) leads to the simple result
\begin{align}
    \frac{\partial g_0}{\partial \Gamma} = -\tilde{\gamma}_f g_0.
\end{align}
where $\tilde{\gamma}_f$, as defined in Eq. (\ref{eqn:gmaster}), is the rate at which "$f$-like" processes occur; these are the processes that lead to the behavior of Eq. (\ref{eqn:betastep}).

This implies that $g_0$ is monotonically decreasing, with $\tilde{\gamma}_f$ controlling the rate. To find an expression for $\tilde{\gamma}_f$ we need to list down the processes that contribute. There are five such processes: (a) Site charge-locking steps of B sites, with rate $b\gamma_f^b$. (b) Site charge-locking steps of R sites that do not lead to a D site creation, with rate $r \gamma_f (1-\gamma_d)$. (c) Site phase-locking of spin sites, with rate $s \gamma_h$. (d) Site phase-locking steps of R sites that create a B site with energy sufficiently high to be immediately decimated, with rate $r\gamma_h (1-k)$; here, $k$ is the probability that a site phase-locking step will result with a B site that has charging energy $<\Omega$; otherwise, the resulting site will be immediately decimated. (e) A variety of bond decimation steps that lead to the creation of a single united site with charging energy so high it is immediately decimated.

Summing all these leads to a flow equation for $g_0$:

\begin{align}
    \frac{\partial g_0}{\partial \Gamma} &= -\left[b \gamma_f^b + r\gamma_f(1-\gamma_d) + s\gamma_h \right.\nonumber \\
    &\left.+ r\gamma_h (1-k) + \left(\cdots\right)g_0\right] g_0 \label{eqn:g0flow}
\end{align}
The inner coefficient of $g_0$ ($\cdots$) is the probability that a bond-decimation step will lead to a subsequent site charge-locking decimation, depending on $f_0$ and $r,b,s$.

The full equations for $s$ and $b$ are still complicated. We shall describe their structure, but postpone presenting the equations themselves to the discussions of the RG fixed points.
The main idea is as follows: Every step either makes a site disappear or changes its type. We go through all these processes, and write down for every process the rate at which it happens and the site types that get created/vanish. Summing all these up, we then renormalize to find the change in the proportions. The most important result of this process is the fact that $\frac{dr}{d\Gamma}<0$: there is no process that creates R sites, and for every process that eliminates other sites -- i.e. site phase-locking steps for D sites, or bond-decimation steps for B sites -- there is an equivalent process that eliminates R sites as well.

This result is manifested in Fig. \ref{fig:heatmap}, where none of the stable phases have R sites (colored in red) -- the red regions are always transients, in the edges of the phase diagram, following a long but finite RG time. It turns out that both $s=1$ (with $r=b=0$) and $b=1$ (with $r=s=0$) can be stable.

In the following subsections, we will discuss two main types of stable phases -- either $g_0 \to 0$ or it stays at some fixed value. For the latter one to happen, we need $f_0 \to 0$, and vice versa -- if $f_0 \not\to 0$, necessarily $g_0 \to 0$, as can be seen from Eq. (\ref{eqn:g0flow}).

A direct equation for $\frac{\partial f_0}{\partial \Gamma}$ can not be written due to its more complicated nature. For $\gamma_h$ we can not write an equation as well, as it is not captured by the ansatz. We will analyze the behavior of these parameters for each phase separately, using appropriate approximations.

\subsection{Superconducting Phase}
We start by discussing the case of finite $g_0$, hence necessarily $f_0=0$ and further $\gamma_h s=0$.
Finite $g_0$ means that the Ansatz is justified -- we will have more and more bond decimations, leading to equations (\ref{eqn:kappastep}) and (\ref{eqn:nstep}), hence the distributions of $\kappa$ and $n$ will be widened. This also means that $\gamma_h$ will be finite, so having $\gamma_h s=0$ requires $s=0$ as well.

\begin{figure}
\includegraphics[width=0.48\textwidth]{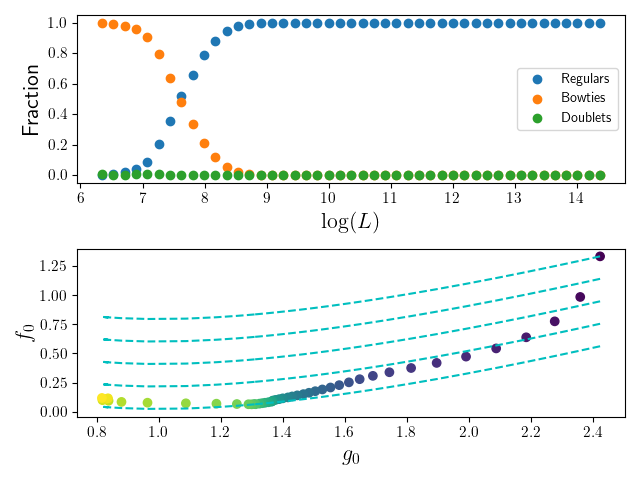}
\caption{
\label{fig:SC}
The flow of the system to the Superconducting point.\\
Top: fractions $r$ (blue), $b$ (orange) and $s$ (green) through the flow; $L$ is the length of the ladder. Bottom: $f_0$ vs $g_0$ through the RG flow (RG time flows from purple to yellow). In light blue, the flow lines in the SC limit, Eq. (\ref{eqn:AKPRepsilon}).
}
\end{figure}

We conclude that a finite $g_0$ stable fixed point is only possible for $b=1$ and $s=0$, in which case we also have $f_0\to0$. This matches the superconducting phases of a single chain\cite{PhysRevB.81.174528}, and we shall not further analyze it. For $f_0\ll1$ the system is a simple chain of superconducting grains, the disorder in $\mu$ has a very weak effect, and the analysis of a system with no chemical potential disorder whatsoever\cite{PhysRevLett.93.150402} holds. This implies in this limit we can expect equation 8 from Ref. [\onlinecite{PhysRevLett.93.150402}] to hold:
\begin{align}
    f_0(\Gamma) = g_0(\Gamma) - \log g_0(\Gamma) - 1 + \varepsilon; \label{eqn:AKPRepsilon}
\end{align}
these flow lines are shown in Fig. \ref{fig:SC}.

\subsection{Insulating Phases}
\label{subsec:InsPhases}
\begin{figure}
\includegraphics[width=0.5\textwidth]{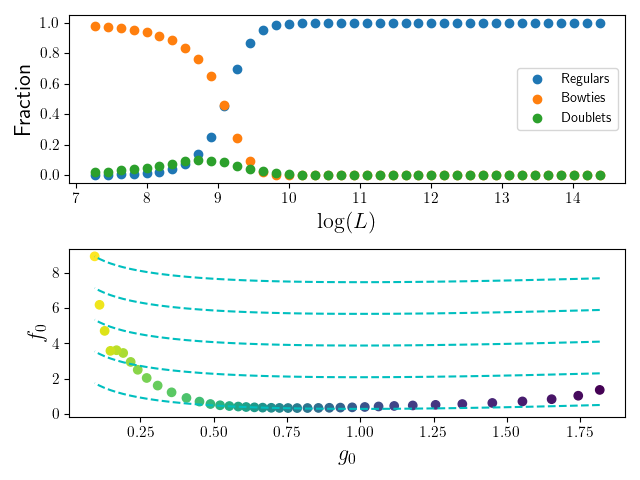}
\caption{
\label{fig:InsB}
The flow of the system to the B-dominated insulating point.\\
Top: fractions $r$ (blue), $b$ (orange) and $s$ (green) through the flow; $L$ is the length of the ladder. Bottom: $f_0$ vs $g_0$ through the RG flow (RG time flows from purple to yellow). In light blue, the flow lines in the SC limit, Eq. (\ref{eqn:AKPRepsilon}).
}
\end{figure}

\begin{figure}
\includegraphics[width=0.5\textwidth]{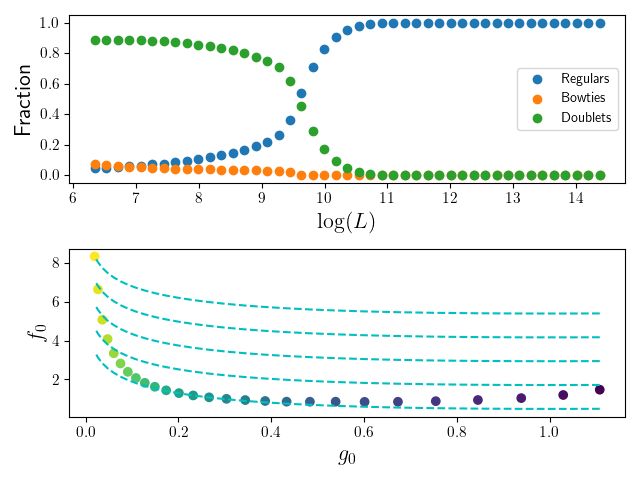}
\caption{
\label{fig:InsD}
The flow of the system to the D-dominated insulating point.\\
Top: fractions $r$ (blue), $b$ (orange) and $s$ (green) through the flow; $L$ is the length of the ladder. Bottom: $f_0$ vs $g_0$ through the RG flow (RG time flows from purple to yellow). In light blue, the flow lines in the SC limit, Eq. (\ref{eqn:AKPRepsilon}).
}
\end{figure}

For $g_0 \to 0$, we get $f_0$ increasing and diverging. While we can not write an analytical equation for $f_0$ to show this directly, this behavior can be observed numerically in figs. \ref{fig:InsB} and \ref{fig:InsD}. Furthermore, in this limit the behavior of $f_0$ is similar to the single-chain case, where it was shown\cite{PhysRevLett.100.170402} to diverge.

In the limit $f_0 \to \infty$ we rarely see bond-decimation steps -- only different types of on-site steps. The distribution of $\zeta$ is very far from a simple exponent, and the distribution of $\bar{n}$ is not uniform at all; instead, the sites are all very close to the $\Delta_i \ll U_i$ regions, which are $\left|n_i\right| \approx \frac{1}{2(1+v)}$ for R rungs and $\left|n_i\right|\approx \frac{1}{2}$ for B rungs, matching Eqs. (\ref{eqn:zeta_n}), (\ref{eqn:fzetan}) and (\ref{eqn:fzetanb}).

In the $f_0\gg 1$ limit we can write the flow equation for $s$:
\begin{align}
\frac{ds}{d\Gamma}&=s\gamma_g\left(\left(1-s\right)^{2}+s^{2}\right)-s\left(1-s\right)\gamma_h\\&+\gamma_f\gamma_dr+\gamma_f(1-\gamma_d)rs+\gamma_f^bsb+sr\gamma_hk
\end{align}
where $k$ is once again the probability that a site phase-locking step of an R site will result in a site charge-locking decimation of the resulting B site, matching Eq. (\ref{eqn:g0flow}).

We recall that we work in the $g_0\to 0$ limit, and that generally $r\to 0$. We hence consider the extreme limit where we substitute $r=\gamma_g=0$ and get a significantly simpler equation:
\begin{align}
\frac{ds}{d\Gamma}&=-sb\gamma_h+\gamma_f^bsb \, .
\label{eqn:dsdgamma}
\end{align}
It is then apparent that there are exactly two processes that matter: site phase-locking steps, dominated by $\gamma_h$ and affecting only D sites; and site charge-locking steps, dominated by $\gamma_f^b$ and affecting only B sites. In this limit the equation for $b$ is necessarily
\begin{align}
\frac{db}{d\Gamma}&=sb\gamma_h-\gamma_f^bsb
\end{align}
as $s+b=1$.

For given constant values of $\gamma_h$ and $\gamma_f^b$, these two equations have two fixed points: $s=0, b=1$ and $s=1, b=0$. Since we start with $s=b=0$ (but with $r=1$) and both types are created through the process, it appears that the only remaining question is the stability -- which point is stable and which is unstable?

The Ansatz leads to a direct result. From subsection \ref{subsec:Ansatz} we have $\gamma_h>1$ and as $g_0\to 0$ we have $\gamma_h\to 2$; meanwhile, Eq. (\ref{eqn:gammafb}) gives $\gamma_f^b<1$ for any finite $f_0$. D sites are decimated faster than B sites and $\frac{ds}{d\Gamma}<0$, so the only stable point is $b=1, s=0$.

However, as can be clearly seen in Fig. \ref{fig:heatmap}, the situation is more complicated than that: there are two different phases in the $g_0\to0$ regime, one B-dominated (blue) and one D-dominated (green). We will discuss this transition in the next subsection, and currently focus on analyzing each phase independently, taking its existence and stability as given.

The B-dominated phase is the Bose Glass mentioned and analyzed in the single-chain system\cite{PhysRevB.81.174528}. The charge degrees of freedom of the islands are frozen, one by one, following very few bond-decimation steps -- hence creating an insulator. Trace amounts of spin sites do not have any significant effect on the flow, since $s$ decays exponentially fast, and because of the second-order coupling terms in Eq. (\ref{eqn:HRDR}), $\tilde{J}_{i-1,i+1}$.

A graph showing the RG flow of the B-dominated insulating phase can be seen in Fig. \ref{fig:InsB}. One can see that the fraction of doublets does initially rise, but they are decimated more quickly than the bowties, hence the phase is Bowtie-dominated. It also shows that $g_0$ does not flow to $0$ very fast, despite monotonically decreasing. Numerically, we can estimate $\gamma_h$ to be $>1.5$ (until there are only B sites left), 
while $\gamma_f^b$ fluctuates around $1$.

Next considering the D-dominated phase corresponding to the fixed point $s=1$ and $b=0$, we see that the Ansatz evidently break down; it turns out that the distribution of $(\zeta, \bar{n})$ does not match Eqs. (\ref{eqn:fzetan}) and (\ref{eqn:fzetanb}).
We can still give careful estimations regarding the nature of this phase. Since $g_0\to 0$, we know that $\beta$ diverges, so $\tilde{J}$ will be very small compared to $\Omega$, while $\kappa$ is evenly distributed. We can therefore expect that the spin chain's behavior will be dominated by the on-site "field" $\vec{K}$ term of Eq. (\ref{eqn:Hspins}), while the inter-site "bond" $\tilde{J}$ term will be irrelevant.

Fig. \ref{fig:InsD} shows the flow towards the D-dominated insulating phase. We can see that there are almost no surviving bowties at all in the flow, only doublets. $f_0$ is diverging until there are no more R and B sites, and the numerics show that $\gamma_h$ for R and D sites are almost identical throughout the flow. When there are bowties,  $\gamma_f^b$ is well above $1$, indicating that the Ansatz definitely does not hold.

\subsection{Spin Glass-Bose Glass Transition}
\label{subsec:SC-BGTrans}
The numerics clearly show that there are two phases with $g_0\to 0$, while the Ansatz indicates only one stable phase. To explain this difference and the evident phase transition, we will have to discuss deviations from the Ansatz. As the initial $g_0$ gets lower and lower, these deviations play a more important role, until they drive a phase transition.

The most important deviation is the distribution of $\bar{n}$. Since the system is initiated with weak disorder in $\bar{n}$, and as Eq. (\ref{eqn:nstep}) only holds when a bond-decimation step is performed, systems with $g_0 \to 0$ might not reach this point.
This is relevant since states with $\bar{n}\approx \frac{1}{2}$ are those responsible for $\gamma_f^b$ not diverging when $f_0 \to \infty$, as they can have infenitesimal $\zeta$, following Eq. (\ref{eqn:zeta_n}). If all the B sites had small $\left|\bar{n}\right|$, their effective vanishing rate $\gamma_f^b$ would be larger.

Another strong effect is the suppression of B-site creation. For $f_0\gg 1$, $\bar{n}\approx \frac{1}{2(1+v)}$ dominates the distribution of R-sites. If such a site undergoes a site phase-locking step, the resulting $\bar{n}$ is around $\frac{2}{2(1+v)}\equiv -\frac{2v}{2(1+v)}$ which is not close to $\pm \frac{1}{2}$ (unless $v\approx 1$). Hence the newly generated sites have very high gap $\Delta$ and are likely to be instantly decimated. To use the terminology of the paragraph before Eq. (\ref{eqn:g0flow}), we might have $k\approx 0$, which means almost no B sites generated.

The last two effects are strongly affected by the fact that $\kappa$ values are initially quite small, and $\bar{n}_0$ quite large. Higher $\kappa$ values lead to B-dominated phase everywhere, with B sites getting generated quickly along the flow and not allowing for D sites to emerge. Lower $\bar{n}_0$ values lead to B sites getting quickly decimated, but it also leads to D sites, that require $\left|\bar{n}\right|>\frac{1}{2(1+v)}$, not getting generated at all.

The numerical results show that if a large amount of B sites are generated, the phase will be dominated by them. However, these results along with the explanations above do not rule out the existence of an exponentially small amount of B sites with $\left|\bar{n}\right|\approx \frac{1}{2}$ among the many D sites. Such a system is expected to be dominated by spin chain physics. One can not rule out the option that an exponentially large RG time would result in a revival of the bowtie fraction. For a detailed discussion of the behavior of long spin chains in such situation, see appendix \ref{app:RDDR}.


The nature of the spin chain phase is dictated by an interplay between a random effective field and random anisotropic interactions, both in the XY plane. The random field appears to be dominant, generating a spin glass state where the spins are randomly polarized. Its behavior is reminiscent of the strongly disordered spin chain models discussed in [\onlinecite{PhysRevB.50.3799}].




\section{Discussion}
\label{sec:Analysis}
We have investigated a two-leg ladder model for a Josephson array, and classified the strong-disorder phases arising from large spatial fluctuations in Josephson ($E_J$) and charging ($E_C$) energies which preserve the $\mathbb{Z}_2$ symmetry relating the two legs of the ladder. This symmetry introduces a spin-like degree of freedom, which plays a key role in the behavior of the system. Following a strong-coupling RG analysis, we identified a phase diagram richer than that of the single Josephson chain, which on a qualitative level confirms the weak-disorder phase diagram obtained in our previous work\cite{PhysRevB.103.245301}.
Specifically, we conclude that the system exhibits three phases: a spin-dominated insulator, which appears to be a spin glass phase; a Bose glass, a disordered charge-localized insulator; and a disordered superconductor with quasi-long-range superconducting correlations. 


The distinction between different phases in the system is expected to be manifested in their different transport behaviors. There are two distinct current channels, and we address transport in each of them as a binary property -- either "insulating" or "superconducting". One channel is symmetric, or longitudinal; in the terms of the scheme depicted in Fig. \ref{fig:Ladder}, it corresponds to the net flow of charge current along the horizontal blue-shaded $J$ bonds. It can be measured in a two-terminal setup where at each end of the ladder both legs are connected to the same contact. The other channel is anti-symmetric, or transverse, corresponding to the net flow of current in the vertical direction, i.e. through the gray $K$ rungs. It can be measured by introducing a short-circuit of the legs at one end of the ladder and connecting the legs to two different contacts at the other end, thus enforcing a counter-flow. We identify a superconducting phase as one that (at zero temperature) perfectly conducts in both channels; an insulating phase is insulating in both channels; and an intermediate phase conducts in one channel and insulates in the other.

The structure of our resulting phase diagram is presented in Fig. \ref{fig:heatmap}. It indicates two phases in which the system can be reduced to a single Josephson chain (blue regions in Fig. \ref{fig:heatmap}). Both phases are superconducting in the anti-symmetric (transverse) channel, as the Josephson couplings $K$ on the rungs are the locally dominant energy scales. However, the two phases are distinct by their longitudinal transport: the disordered superconductor (SC in Fig. \ref{fig:heatmap}) where $E_J$ is dominant is superconducting in both channels, while the Bose glass phase (denoted by BG in the figure), where $E_C$ dominates, is insulating. The latter is identified as an \textit{intermediate phase}, since it is superconducting to transverse current.


The third phase (green region in Fig. \ref{fig:heatmap}) is predicted to be a full-fledged insulator of a glassy nature. Its behavior is best described by the spin-like degree of freedom on the rungs, which manifest a spin-glass (SG) state where the spins are polarized in random directions in the XY-plane. In terms of the original Josephson-ladder degrees of freedom, this phase corresponds to an effective random flux state where the net Josephson coupling (and hence conduction of super-current) in the transverse direction is suppressed. At the same time, the spin-spin coupling terms of Eq. (\ref{eqn:HDD}) flow to $0$ rapidly. The system is rather dominated by the on-site energy scales, which imply a charge insulator in the longitudinal direction as well.

We conclude that the two-leg ladder structure, even in the strongly disordered limit, supports an intermediate phase separating the two quantum phases characteristic of its single-chain analogue\cite{PhysRevLett.93.150402}. This demonstrate that the addition of legs serves as a gateway to a dimensional crossover to a two-dimensional system, staying tractable and yet capturing certain aspects of the interplay between disorder and interactions in dimensions greater than one.
We suggest that further analysis of these phases can be done in various directions and shed more light on the nature of the phase transitions. Numerical tools have previously been shown to be useful in analyzing disordered phase transitions\cite{PhysRevB.94.134501}, showing evidence for intermediate phases in various different systems\cite{PhysRevB.111.094212}. 

A possible realization of our investigated systems is an engineered Josephson ladder. When put in the strong quantum fluctuations regime (with typical $E_J\sim E_C$) and subjected to symmetry-preserving disorder, we predict that such a ladder will undergo an indirect SIT with an intermediate phase. Apart from transport measurements, we expect the distinction between the phases to be signified by measurements of thermodynamic quantities, e.g. the superconducting stiffness\cite{GAVISH2021132767, Keren_2022}.

As a final remark, we note that the restriction of perfect $\mathbb{Z}_2$  symmetry requires an accurate control of the Josephson array parameters, which is notoriously challenging to achieve in realistic devices. We expect that a tiny violation of the symmetry between the coupled legs might still allow observation of the key phenomena; however, the true ($T \to 0$) quantum state will reduce to a single-chain behavior. As an alternative realization of our model, one can consider a system with a different type of discrete degree of freedom which does not correspond to spatial separation between two perfect duplicates of the single chain. A cold-atom system,  composed of unit cells characterized by some internal degree of freedom (e.g. the level index of a Rydberg atom or spin of an atom), could serve as a good example. Such a system may prove easier to control in a way that preserves the perfect $\mathbb{Z}_2$ symmetry; previous works have already used optical lattices to demonstrate that the interplay of disorder and interactions can lead to various phase transitions\cite{Yu2024, PhysRevB.110.184202}.

\begin{acknowledgements}
    We gratefully acknowledge useful discussions with Ehud Altman, Ganpathy Murthy, Jonathan Ruhman, Thomas Vojta, Gil Refael and Atanu Rajak. E. S. thanks the Aspen Center for Physics (NSF Grant No. 1066293) for its hospitality. This work was supported by the Israel Science Foundation (ISF) Grant No. 993/19.
\end{acknowledgements}

\appendix

\section{Bond Decimations Derivations}
\label{app:Bonds}

In this appendix we shall derive Eqs. (\ref{eqn:Jdecim1})-(\ref{eqn:Jdecim4}) that describe bond decimations, and other results from the main text that regard the dynamics of a locally dominant phase coupling term.
To derive these results, one calculates the Lagrangian of the system, which allows to find the equations that describe bond decimations, as well as site phase-locking steps.

We start with the charge sector of equations (\ref{eqn:Hsum})-(\ref{eqn:Hrung}). Since the system will be dominated by $J$ terms, we will substitute $\phi_{\nu,i}=\phi_{\nu,i+1}$ and $\dot{\phi}_{\nu,i}=\dot{\phi}_{\nu,i+1}$ into the Lagrangian. To derive the Lagrangian, we start with the Hamiltonian of the charge sector:
\begin{align}
    \mathcal{H}_i^C = \frac{1}{2} U_i^2 \left(n_{u i}^2 + n_{di}^2\right) + V n_{u i} n_{d i} - \mu_i \left(n_{u i} + n_{d i}\right) \label{appC:Hc}
\end{align}
and since $\phi, n$ are canonical conjugates with
\begin{align}
    \left[\phi_{\nu i}, n_{\nu^\prime j}\right]=\delta_{i,j}\delta_{\nu,\nu^\prime}
\end{align}
we can write the equations of motion:
\begin{align}
    \dot{\phi}_{\nu i}=\frac{\partial H}{\partial n_{\nu i}}=U_i n_{\nu i}-\mu_i +V_i n_{-\nu i}
\end{align}
where $-\nu$ is the side opposite of $\nu$.

We extract $n_{\nu i}$:

\begin{align}
    n_{\nu i} = \frac{U_i}{U_i^2-V_i^2} \dot{\phi}_{\nu i} - \frac{V}{U_i^2-V_i^2} \dot{\phi}_{-\nu i} + \frac{\mu_i}{U_i+V_i}.
\end{align}

Now we write the charge part of the Lagrangian:
\begin{align}
    L_i^C &= n_{u i} \dot{\phi}_{u i} + n_{d i} \dot{\phi}_{d i} - H_i^C \\
     &= \frac{1}{2}\frac{U_i}{U_i^2-V_i^2}\left(\dot{\phi}_{u i}^2 + \dot{\phi}_{d i}^2\right) - \frac{V_i}{U_i^2-V_i^2} \dot{\phi}_{u i} \dot{\phi}_{d i} \nonumber \\ &+ \frac{\mu_i}{U_i+V_i} \left(\dot{\phi}_{u i} + \dot{\phi}_{d i}\right) + Const. \label{appC:RLagrangian}
\end{align}

Based on this result, we now picture two copies of this local Lagrangian in two adjacent sites, indexed $i$ and $i+1$, connected by strong phase bonds 
\begin{align}
    -J \left[\cos(\phi_{u i} - \phi_{u i+1}) + \cos(\phi_{d i} - \phi_{d i+1})\right]
\end{align}
and these strong bonds are equivalent to setting $\phi_{u i} = \phi_{u i+1}$. Since there is no charge coupling between adjacent unit cells, we sum the coefficients of matching terms. For example, if we look on the coefficient of $\dot{\phi}_{u i}^2 = \dot{\phi}_{u i+1}^2$, it will be the sum of the two matching coefficients of $L_i^C$ and $L_{i+1}^C$:
\begin{align}
    \frac{1}{2} \frac{U_1}{U_1^2-V_1} + \frac{1}{2} \frac{U_2}{U_2^2-V_2^2}
\end{align}
and since the structure is kept, the structure of the Hamiltonian is the structure of Eq. (\ref{appC:Hc}), and the expression above is equal to $\frac{1}{2}\frac{U_{eff}}{U_{eff}^2-V_{eff}^2}$. The equations for the quadratic coefficients (of $\dot{\phi}_{u/d}^2$ and $\dot{\phi}_u\dot{\phi}_d$) are
\begin{align}
    \frac{1}{2}\frac{U_{eff}}{U_{eff}^2-V_{eff}^2} = \frac{1}{2} \frac{U_1}{U_1^2-V_1} + \frac{1}{2} \frac{U_2}{U_2^2-V_2^2} \\
    \frac{1}{2}\frac{V_{eff}}{U_{eff}^2-V_{eff}^2} = \frac{1}{2} \frac{V_1}{U_1^2-V_1} + \frac{1}{2} \frac{V_2}{U_2^2-V_2^2}.
\end{align}

Summing the two leads to Eq. (\ref{eqn:Jdecim4}) and substructing leads to Eq. (\ref{eqn:Jdecim3}).

Explicitly solving these equations is not illuminating. We get
\begin{align}
    U_{eff} =& \frac{U_1^2U_2 + U_2^2U_1 - U_2V_1^2 - U_1V_2^2}{(U_1+U_2)^2-(V_1+V_2)^2} \\
    V_{eff} =& \frac{U_1^2V_2 + U_2^2 V_1 - V_1^2 V_2 - V_2^2 V_1}{(U_1+U_2)^2 - (V_1+V_2)^2}
\end{align}

These results are not very useful if one is looking to understand the behavior of $U_{eff}$, but they are important to determine the dynamics of $v=\frac{V}{U}$. Let $v_{1,2,eff}=\frac{V_{1,2,eff}}{U_{1,2,eff}}$ and we get

\begin{align}
    v_{eff} = \frac{U_1v_2(1-v_1^2)+U_2v_1(1-v_2^2)}{U_1(1-v_1^2) + U_2(1-v_2^2)} \label{appeq:veff}
\end{align}
and this is a weighted average of $v_1$ and $v_2$ with weights $U_1(1-v_1^2)$ and $U_2(1-v_2^2)$ correspondingly -- justifying Eq. (\ref{eqn:vAssumption}).

The analysis for the linear term is simpler -- the coefficient of $\dot{\phi}_{u/d\ i}$ is $\frac{\mu_i}{U_i+V_i} \equiv \bar{n}_i$, so merging two adjacent sites leads directly to Eq. (\ref{eqn:Jdecim2}).

The remaining equation, (\ref{eqn:Jdecim1}), is even simpler -- this is not a charge term, so it is the same in the Hamiltonian and the Lagrangian, and we get

\begin{align}
    K_{eff} \cos\left(\phi_{u}-\phi_{d} - \varphi_{K_{eff}}\right) =& K_1 \cos\left(\phi_{u}-\phi_{d} - \varphi_{K_1}\right) \nonumber \\ +& K_2 \cos\left(\phi_{u}-\phi_{d} - \varphi_{K_2}\right)
\end{align}

and Eq. (\ref{eqn:Jdecim1}) is a direct result.

This concludes case of bond decimation between two R sites. What about a phase-locking step, creating a B site? A site phase-locking step means letting $\phi_{u i}=\phi_{d i}$. We start by writing the charge part of the Hamiltonian of a B site, Eq. (\ref{eqn:HB}), and its Lagrangian is straightforward:

\begin{align}
    \dot{\phi}_i =& \frac{\partial H_B^i}{\partial n_i} = 2 U^B_i n_i - \mu_i \\
    n_i =& \frac{\dot{\phi}_i}{2U^B_i} + \frac{\mu_i}{2U^B_i} \\
    L_B^i =& \frac{\dot{\phi}_i^2}{4U^B_i} + \frac{\mu_i}{2U^B_i} \dot{\phi}_i.
\end{align}

Letting $\dot{\phi}_{ui} = \dot{\phi}_{di} \equiv \dot{\phi}_i$ in Eq. (\ref{appC:RLagrangian}) gives

\begin{align}
    L=\frac{1}{2(U_i+V_i)} \dot{\phi}_i^2 + \frac{2\mu_i}{U_i+V_i} \dot{\phi}_i
\end{align}

which immediately gives $U^B_i=\frac{U_i+V_i}{2}$ and $\mu_i^{eff}=2\mu_i$.

A bond decimation step of two B sites is simple, and it appears in the case of a single chain\cite{PhysRevLett.93.150402} -- similar to the bond-decimation step of the regular rungs, we let $\phi_i = \phi_{i+1}$ to get
\begin{align}
    \frac{1}{4U^B_{eff}} = \frac{1}{4U^B_i} + \frac{1}{4U^B_{i+1}} \\
    \frac{\mu_{eff}}{2U^B_{eff}} = \frac{\mu_i}{2U^B_i} + \frac{\mu_{i+1}}{2U^B_{i+1}}.
\end{align}

A bond decimation of an $R_b$ or a $B_r$ means to start with $L_R^i + L_B^{i+1}$, and set $\phi_{u i} = \phi_{d i} = \phi_{i+1}$. To do so, we can first let $\phi_{u i} = \phi_{d i}$, get the effective B site, then join them together like B sites.

When D sites are involved, the situation is simpler. Since their charge gap is very large, we assume that $U_2, V_2, \mu_2$ of the D site are significantly larger than $U_1, V_1, \mu_1$ of the R site and take the limit $U_2\to \infty$ while keeping $\frac{\mu}{U}, \frac{V}{U}$ constant we can simply write $U_{eff}=U_1$, $V_{eff}=V_1$. For $\mu$ the situation is more complicated as $\bar{n}$ is additive; indeed, one of the parameters of a spin site is its original $\bar{n}$, and when a bond-decimation step is performed, Eq. (\ref{eqn:Jdecim2}) still holds.

A bond-decimation step of D sites results with a new site that has very large $U, V$ but $\bar{n}$ that has only one non-degenerate ground-state, hence both sites diappear, similar to singlet sites in [\onlinecite{PhysRevB.81.174528}].

\subsection*{A Comment about the definition of $U^B$}
The definition of $U^B_i$ in Eq. (\ref{eqn:HB}) is inconsistent with the charge sector of each rung in Eq. (\ref{eqn:Hleg}). This is due to a more global selection of non-universal coefficient that slightly simplifies the numerical simulations.

When an $R_b$ site undergoes an on-site phase decimation and turns into a $B_b$ site, the coupling $J$ appears to double, as two parallel bonds turn into one. To avoid that, we define $J$ not as the value of every single connection, but rather as half of the total value of the $J$ terms between two adjacent sites. That means all the following have the same $J$ when we perform statistics on $J$ values:

\begin{align}
    &-J \left[\cos(\phi_{u,i} - \phi_{u, i+1}) + \cos(\phi_{d, i} - \phi_{d, i+1})\right] \\
    &-\frac{J}{2} \left[\cos(\phi_{u,i} - \phi_{u, i+1}) + \cos(\phi_{d, i} - \phi_{d, i+1}) + \right. \nonumber \\ &\ \left. \cos(\phi_{u,i} - \phi_{d, i+1}) + \cos(\phi_{d, i} - \phi_{u, i+1})\right] \\
    &-2J\left[\cos(\phi_i - \phi_{i+1})\right]
\end{align}
representing $R_{r=}, R_{r\times}$ and $B_b$ sites, respectively; as well as $R_b$ and $B_r$ sites (very similar to $R_{r=}$ case).

For the $V=\mu=0$ case, a site charge decimation of an $R_r$ site leads to $J_{eff}=\frac{2J_1J_2}{U}$, as $\Delta=\frac{U}{2}$. For simplicity we would like to have the equivalent equation to be true for $B_b$ sites, $J_{eff} = \frac{2J_1J_2}{U^B}$. For this to be the case with the moderated $J$ definition, a factor of $2$ in the definition of $U^B$ is required.

\section{Schrieffer-Wolff Transformations}
\label{app:S-W}
In this appendix we will derive the calculations leading to equations (\ref{eqn:Udecim}) and (\ref{eqn:RDterm}) using Schrieffer-Wolff transformations. We will derive both the structure of the results and the numerical coefficients; for each coefficient we will have to test both its behavior in the general case (for generic values of $\mu$) and in the specific case of $f_0 \gg 1$ where $\left|\bar{n}\right|\approx \frac{1}{2}$, to make sure that no term vanishes in either case.

The leading idea of Schrieffer-Wolff transformations is as follows. We have a system split into two subspaces, low-energy and high-energy. We'd like to write an effective description for the low-energy sector, that doesn't couple to the high-energy sector. We will denote the projection to the low(high)-energy subspace $\mathcal{P}_{0(1)}$, so $\mathcal{P}_0+\mathcal{P}_1=Id$. We will categorize terms to "on-diagonal", coupling a sector (mainly the low-energy sector) to itself, and "off-diagonal", coupling between the sectors.

In our case, the system is gapped -- we write
\begin{align}
\mathcal{H}=\mathcal{H}_0 + V,
\end{align}
where $\mathcal{H}_0$ is block-diagonal and gapped and $V$ is neither (in the most general case). We would like to work in the large-gap limit, where $V\ll\Delta$, and we will work perturbatively. Following our base Hamiltonian in Eqs. (\ref{eqn:Hsum})-(\ref{eqn:Hrung}), we will focus on site charge-locking steps, where $\mathcal{H}_0$ is the charge term including $U_i,V_i,\mu_i$ and $V$ will be all the phase terms that include the sites of rung $i$. All other terms will not be affected from our analysis and are kept unchanged.

The Shcrieffer-Wolff transformation is a change of variables for the sites involved, which we denote by $e^S$ (where $S$ is an operator). We are looking for $S$ such that $e^{-S}(\mathcal{H}_0+V)e^S$ has no off-diagonal terms, in the sense that it decouples the high-energy and low-energy sectors. As we work perturbatively, we assume that $V$ and $S$ are both small and expand to second order:

\begin{align}
e^S \mathcal{H} e^{-S} &= \mathcal{H}_0 + (V + [S,\mathcal{H}_0]) \\ &+ ([S,V] + \frac{1}{2} [S,[S,\mathcal{H}_0]]) \nonumber
\end{align}.

We shall choose $S$ such that $V=-[S,\mathcal{H}_0]$, to cancel out the leading order and get a second-order correction $-\frac{1}{2}[S,[S,\mathcal{H}_0]]$. Canceling out the off-diagonal terms of the leading order, the most important interaction of the low-energy sector will be the second-order on-diagonal term, which is given by

\begin{align}
    \mathcal{H}^{(2)} = \mathcal{P}_0 [S, [S, \mathcal{H}_0]] \mathcal{P}_0 = -\frac{1}{2} \mathcal{P}_0 [S, V] \mathcal{P}_0
\end{align}

Still in the general case, we now introduce notations for the eigen-basis of $\mathcal{H}_0$. It has two sets of vectors - $\ket{n0}$ for the low-energy sector and $\ket{m1}$ for the high-energy sector:

\begin{align}
    \mathcal{H}_0=& \sum_n E_{n0} \left|n0\right\rangle\left\langle n0\right| + \sum_m E_{m1} \left|m1\right\rangle\left\langle m1\right| \\
    V_{od} =& \sum_{n,m} V_{nm} \left|n0\right\rangle\left\langle m1\right| + h.c. \\
    S =& \sum_{n,m} S^{01}_{nm} \left|n0\right\rangle\left\langle m1\right| + S^{10}_{nm} \left|m1\right\rangle\left\langle n0\right|
\end{align}
in our case, we will take the charge basis for the two sectors. Note that we assume $V$ doesn't have on-diagonal terms - if it does, we consider them part of $\mathcal{H}_0$. For $V$ to be hermitian, we need $S$ to be anti-hermitian, $S^{10}_{nm}=-S^{01}_{nm}$; we let $S^{01}_{nm} \equiv S_{nm}$ (hence $S^{10}_{nm} = -S_{nm}$), and get

\begin{align}
    S_{nm} = \frac{V_{nm}}{E_{n0}-E_{m1}}.
\end{align}

This leads to the overall correction

\begin{align}
    \frac{1}{2} \mathcal{P}_0 [S,V] \mathcal{P}_0 =& \sum_{nk} \left(\sum_m \frac{V_{nm} V_{km}}{2} \cdot \alpha_{nmk}\right) \left|n0\right\rangle \left\langle k0 \right| \label{appeq:SWterm} \\
    \alpha_{nmk} =& \left(\frac{1}{E_{n0}-E_{m1}}+\frac{1}{E_{k0}-E_{m1}}\right) \nonumber
\end{align}

To keep track in our specific case, it is useful to get a basic understanding of this general result. For every two states in the low-energy subspace $n$ and $k$, we sum over all the possible states in the high-energy sector $m$, and take into account the second-order terms that "go through" $m$ as an intermediate step. Such processes are second-order in $V$, hence they depend on the multiplication of the matrix elements $V_{nm}V_{km}$. Such process is "damped" by the energy gap between $m$ and either $n$ or $k$ - which is assumed to be significantly larger than $V$.

\subsection{The Non-Degenerate Case}
We shall start by describing the case of site charge-locking step where the site is totally decimated, as the ground-state is not degenerate at all.

For simplicity, we will use the indices $0, 1$ and $2$, where site $1$ is the decimated site, and $0, 2$ are its neighbors. $J_{0(1)}$ are the bonds to the left (right). $U, V, \mu$ are all the parameters of site $1$.

To turn phase operators into the charge basis, we reacll the bosonization $b \propto e^{i\phi}$ and $b^\dagger \propto e^{-i\phi}$, so $\cos(\phi_{i\nu}-\phi_{j\nu^\prime}) \propto b^\dagger_{i\nu} b_{j\nu} + h.c.$ -- a hopping term.

In our system, for $\left|\mu\right|<\frac{U}{2}$, the charge ground state is not degenerated, hence $\mathcal{P}_0$ is just projection on the $n_{u,1}=n_{d,1}=0$ subspace. Note we only project to this subspace in a specific unit cell, and the rest change freely. Restricting ourselves that way, there are only transitions with $\left|0n\right\rangle=\left|0k\right\rangle$, or transitions that pass one charge "through" the decimated site, creating a next-nearest-neighbor term. We calculate using Eq. (\ref{appeq:SWterm}), substituting $E_{n0}$ and $E_{m1}$ that are only affected from the Hamiltonian of the decimated site $1$, and get the coefficient
\begin{align}
    J_{0, 2}^{eff} = J_0 J_1 \left(\frac{1}{\frac{1}{2}U-\mu} + \frac{1}{\frac{1}{2}U+\mu}\right)
\end{align}
using $\frac{1}{2}U-\mu=\Delta$ and $\frac{\mu}{U}=\bar{n}(1+v)$ we can re-write this as
\begin{align}
    J_{0, 2}^{eff} = \frac{J_0J_1}{\Delta} \left(1+ \frac{1-(1+v)\left|\bar{n}\right|}{1+(1+v)\left|\bar{n}\right|}\right) \label{appeq:Jeff}
\end{align}
and this matches the result of the single chain case\cite{PhysRevB.81.174528}. This is also identical to Eq. (\ref{eqn:Udecim}) up to a numerical pre-factor between $1$ and $2$.

\subsection{The 2-fold Degenerate Case}
There are three main differences for the case of $\left|\mu\right|>\frac{U}{2}$. The first is that the subspace we are limited to is larger, with two states, not just one. The second is that there are more terms, including $\tilde{J}_{0, 2, \times}$ of Eq. (\ref{eqn:HRDR}). The third is that the magnitude of the terms might depend on the internal state of the resulting spin in site 1 -- so the coefficient of terms like $\cos(\phi_{u0}-\phi_{u2})$ might not be equal to the coefficient of its matching up-down symmetric term, $\cos(\phi_{d0}-\phi_{d2})$.

Combining everything together and neglecting the terms that are the unit matrix, we get five terms:
\begin{align}
    -J_{02}^{NNN} \left[\cos(\phi_{0u}-\phi_{2u}) + u.d.c.\right] \label{appeq:Js1}\\
    -J_{02,as}^{NNN} \left[\cos(\phi_{0u}-\phi_{2u}) - u.d.c.\right] \sigma_1^z \\
    -\tilde{J}_{01} \left[\sigma_1^+ b_{0d}^\dagger b_{0u} + h.c.\right] \\
    -\tilde{J}_{12} \left[\sigma_1^+ b_{2d}^\dagger b_{2u} + h.c.\right] \\
    -J_{02,\times}^{NNN} \left[\sigma_1^+b_{2d}^\dagger b_{0u} + h.c.\right] \label{appeq:Js5}
\end{align}
where each of the coefficients matches a selection of $n, k$ in Eq. (\ref{appeq:SWterm}).

We can calculate these coefficients one by one. Like in the non-degenerate case, the energies $E_{n0}$ or $E_{m1}$ can be approximated to depend only on $n_{1u}, n_{1d}$, so we write $E_{\ket{n_un_d}}$ to mean the energy for $n_{u1}=n_u$ and $n_{d1}=n_d$. For the next-nearest-neighbors we get:
\begin{widetext}
\begin{align}
    J_{02}^{NNN} =& \frac{J_0J_1}{2} \left[\frac{1}{E_{\ket{-11}}-E_{\ket{01}}} + \frac{1}{E_{\ket{11}}-E_{\ket{01}}} + \frac{1}{E_{\ket{00}}-E_{\ket{01}}} + \frac{1}{E_{\ket{02}}-E_{\ket{01}}}\right] \\
    =& \frac{J_0J_1}{2} \left[\frac{U}{\frac{U^2}{4}-(\mu-V)^2} + \frac{U}{\frac{U^2}{4}-(\mu-U)^2}\right]
    \\
    J_{02, as}^{NNN} =& \frac{J_0J_1}{2} \left[\frac{1}{E_{\ket{-11}}-E_{\ket{01}}} + \frac{1}{E_{\ket{11}}-E_{\ket{01}}} - \frac{1}{E_{\ket{00}}-E_{\ket{01}}} - \frac{1}{E_{\ket{02}}-E_{\ket{01}}}\right]\\
    =& \frac{J_0J_1}{2} \left[\frac{U}{\frac{U^2}{4}-(\mu-V)^2} - \frac{U}{\frac{U^2}{4}-(\mu-U)^2}\right] \\
    J_{02, \times}^{NNN} =& \frac{J_0J_1}{2} \left[\frac{1}{\frac{1}{2}U+V-\mu} + \frac{1}{\mu-\frac{1}{2}U}\right]
\end{align}
\end{widetext}
Each of these terms has a different nonuniversal coefficient, but they are all $O(\frac{J_1J_2}{U})$. If $\Delta=\mu-\frac{U}{2} \ll U$, all these terms will be dominated by $ \frac{J_0J_1}{2}\frac{1}{\mu-\frac{U}{2}}=\frac{J_1J_2}{2\Delta}$. That means in the $\Delta \ll U$ limit, matching $f_0 \gg 1$, no term vanishes in this limit.

For the nearest-neighbors terms, we have
\begin{widetext}
\begin{align}
    \tilde{J}_{01} = \frac{J_0^2}{2} \left[\frac{1}{E_{\ket{11}}-E_{\ket{10}}} + \frac{1}{E_{\ket{00}}-E_{\ket{10}}}\right]=\frac{J_0^2}{2} \left[\frac{1}{\frac{1}{2}U+V-\mu} + \frac{1}{\mu-\frac{1}{2}U}\right]
\end{align}
\end{widetext}
and similarly for $\tilde{J}_{12}$.
Here as well, the general behavior is $O(\frac{J^2}{U})$, but if $\Delta\ll U$ there is a $\frac{J^2}{2\Delta}$ term - so both in the $f_0 \gg 1$ limit and in the general case, the we have $J^{NNN} \sim \sqrt{\tilde{J}_{01}\tilde{J}_{12}}$.

Like we mentioned just before Eq. (\ref{eqn:HDD}), if the adjacent site undergoes a charge-locking step we do not have to apply second-order Schrieffer-Wolff once again, since $\mathcal{P}_0V\mathcal{P}_0 \neq 0$ and it is the dominant term. This term is given in Eq. (\ref{eqn:HDD}). The exact behavior of long spin chains is discussed in appendix \ref{app:RDDR}.

It should be noted that, despite being allowed by symmetry, there is no $\sigma_z\sigma_z$ term in the Hamiltonian. Careful analysis shows that if site $i$ undergoes a charge-locking step followed by a charge-locking step in site $i+1$, the coefficient of $\sigma^z_i\sigma^z_{i+1}$ is smaller by a factor of $\frac{\Delta_{i+1}}{\Delta_{i}}$, the ratio of the gaps of the sites; and the strong-disorder limit means this ratio is large.

If the original site was $R_{r\times}$ (and then $D_{r\times}$ or $R_{d\times}$), the up-down behavior is symmetrized -- the cross-connections allow for terms that are identical to (\ref{appeq:Js1})-(\ref{appeq:Js5}) but with $b_{0u}$ replaced with $b_{0d}$ and $b_{0u}^\dagger$ with $b_{0d}^\dagger$. This also doesn't allow for $\sigma_z\sigma_z$ interactions at all. If the R side of such connection undergoes a site charge-locking step and turns into a spin site, we call the resulting site $D_{d\times}$, as can be seen in Fig. \ref{fig:DRungs}.

The difference between $D_{r\times}$ and $D_r$ sites, and similarly between $D_{d\times}$ and $D_d$ sites, does not appear important in our ladder framework; it might, however, play a more important role in a thorough analysis of the spin chain phase.

We conclude that Eq. (\ref{eqn:Udecim}) and (\ref{eqn:RDterm}) are true up to a non-universal constant of order unity, both if $U\approx \Delta$ and if $\Delta \ll U$; and we ignore said non-universal constant in our analysis of the flow, both numerically and analytically.

\section{Long Spin Chains Analysis}
\label{app:RDDR}

In the main text, we have discussed the behavior of two adjacent doublet sites and their interaction in Eq. (\ref{eqn:HDD}).
However, if these two sites were to be decimated, the system's behavior will be dominated by long-range terms -- next-next-nearest neighbors between the two R sites surrounding the D sites. Moreover, the nearest-neighbor terms here do not allow for symmetric current to flow at all, so the next-next-nearest neighbors terms dictate the behavior of the longitudinal conductance.

Similar to this situation, which we will schematically denote with R-D-D-R, we can hold a similar discussion for longer spin chains between R and B sites, like R-D-D-D-B. In this appendix we will focus on R-D-D-R chains and show the approximations and inaccuracies in their description throughout the main text. Once longer chains appear, we are in the spin-dominated phase and the ladder-based analysis is no longer complete.

First, recall the situation for a single spin site between two R sites. There are two nearest-neighbors connections, $\tilde{J}_{01}$ and $\tilde{J}_{12}$. However, in some situations we are concerned by the effective second-order next-nearest-neighbors term $J_{02}^{NNN}$. It turns out that $J_{02}^{NNN} = \sqrt{\tilde{J}_{01} \tilde{J}_{12}}$, hence we can keep the strengths of the local NN connections, and "restore" the value of $J_{02}^{NNN}$. As we will see, this will not be the case for an R-D-D-R chain with two spin sites between the regular rungs -- the information in each separate unit cell will not suffice to restore the strengths of the long-range couplings.

To discuss an R-D-D-R chain we start with a small chain of four sites, all regular rungs, $0, 1, 2, 3$, and assume that $1$ undergoes a charge-locking step, creating a D site with charging energy $\Delta_1=\Omega_1$. We further assume that later on, $2$ undergoes a charge-locking step, creating another spin site with charging energy $\Delta_2=\Omega_2$. We will assume that any other energy scale in the system is $\ll \Omega_2$ and that $\Omega_2 \ll \Omega_1$, and find the types of interaction and their strengths. We will use $J_0, J_1, J_2$ and $\Omega_1, \Omega_2$. Note that the unit cells of this chain is currently $R_dD_dD_rR_*$ (where the type of the last R site is unknown); the behavior will be identical if $R_d$ will be replaced with $R_{d\times}$ or $B_d$, if $D_d$ will be replaced with $D_{d\times}$ and if $D_r$ will be replaced with $D_{r\times}$ or $D_b$.

The nearest-neighbor terms are quite simple:
\begin{align}
    \tilde{J}_{01} = \frac{J_0^2}{\Omega_1} \\
    \tilde{J}_{12} = \frac{J_1^2}{\Omega_1} \\
    \tilde{J}_{23} = \frac{J_2^2}{\Omega_2}.
\end{align}

Note that $\tilde{J}_{12}$ is not symmetric between $\Omega_1$ and $\Omega_2$.

There are also next-nearest terms. After site $1$ was decimated and before site $2$ was decimated, there were connections $J_{02}^{NNN}=\frac{J_0J_1}{\Omega_1}$, hence after the decimation of site $2$ we have
\begin{align}
    \tilde{J}_{02} = \frac{J_0^2J_1^2}{\Omega_1^2\Omega_2}.
\end{align}

As to sites $1$ and $3$, their connection is created from a process like the one described in Eq. (\ref{eqn:UdecimJD}):
\begin{align}
    \tilde{J}_{13} = \frac{J_1^2 J_2^2}{\Omega_1 \Omega_2^2}.
\end{align}

Both these next-nearest neighbors terms are fourth-order in the original $J$s, hence they will be negligible compared to the terms described next.

The connection between sites $0$ and $3$ is different, as it maintains the original structure of the phase Hamiltonian in Eq. (\ref{eqn:Hleg}).
Before site $2$ was decimated we have $J_{02}^{NNN}=\frac{J_0J_1}{\Omega_1}$ and $J_{23}$, so after site 2 is decimated we have a next-next-nearest-neighbors term with coefficient

\begin{align}
    J_{03}^{NNNN} = \frac{J_{02}^{NNN}J_{2}}{\Omega_2}=\frac{J_0J_1J_2}{\Omega_1\Omega_2}.
\end{align}

This seems simple enough -- a simple, symmetric expression that can be written in terms of $J$ and $\Omega$. The problem is that through the flow the unit cells change, and $\Omega$s are no longer parameters of these cells -- only the local $\tilde{J}_{i,i+1}$ coefficients. It can be easily seen that $J_{03}$ cannot be written with the three of them -- one can only write
\begin{align}
    J_{03}^{NNNN} = \frac{\sqrt{\tilde{J}_{01}\tilde{J}_{12}\tilde{J}_{23}}}{\sqrt{\Omega_2}}. \label{appB:J03eff}
\end{align}

Furthermore, assume we now decimate site $1$ in a site phase-locking step with $\Omega_3\ll\Omega_2$. The interaction between sites $0$ and $2$ will have two contributions -- the existing $\tilde{J}_{02}$ and a new second-order contribution
\begin{align}
    \tilde{J}_{02}^\prime = \frac{\tilde{J}_{01}\tilde{J}_{12}}{\Omega_3}=\frac{J_0^2J_1^2}{\Omega_1^2\Omega_3} \gg \tilde{J}_{02}
\end{align}
so the new second-order contribution will dominate. The resulting chain will have $\tilde{J}_{02}^{eff}=\frac{J_0^2J_1^2}{\Omega_1^2\Omega_3}$ and $\tilde{J}_{23}$ like the original chain. This justifies the earlier claim that $\tilde{J}_{02}$ is negligible.

Suppose we further decimate site 2. We will deduce that the effective connection is now
\begin{align}
    J_{03}^{deduced}=&\sqrt{\tilde{J}_{02}^{eff}\tilde{J}_{23}}=\frac{J_0J_1J_2}{\Omega_1\sqrt{\Omega_2\Omega_3}} \\ =&\sqrt{\frac{\Omega_2}{\Omega_3}}J_{03}^{NNNN}\gg J_{03}^{NNNN}
\end{align}

We stress that this is a fundamental problem -- the information required to determine the value of $J_{03}^{NNNN}$ does not exist within the local terms $\tilde{J}_{i,i+1}$. There is no way to combine the local information into the correct value $J_{03}^{NNNN}$ without knowing $\Omega_i$. In the absence of a better solution, we use Eq. (\ref{appB:J03eff}) with $\Omega$ instead of $\Omega_2$, hence missing a factor of $\sqrt{\frac{\Omega}{\Omega_2}}$. Our estimations for $J$ terms in phases with long spin chains are therefore too high.

A third scenario to consider is a bond decimation of $\tilde{J}_{01}=\Omega_3$. In such case the connection between the new site (we will index it with $0$) and site $2$ will be $\tilde{J}_{12}$, so we will deduce
\begin{align}
    J_{03}^{deduced} = \sqrt{\tilde{J}_{12}\tilde{J}_{23}}=\frac{J_1J_2}{\sqrt{\Omega_1\Omega_2}}.
\end{align}

From the definition of $\Omega_3$ we have $J_0=\sqrt{\Omega_1\Omega_3}$, so we once again get
\begin{align}
    J_{03}^{deduce} = \frac{J_0J_1J_2}{\Omega_1 \sqrt{\Omega_2\Omega_3}} = \sqrt{\frac{\Omega_2}{\Omega_3}} J_{03}^{NNNN}.
\end{align}

Once again, we use the wrong value of $\Omega$ when estimating $J_{03}$ and get an optimistic result by a factor of $\sqrt{\frac{\Omega_{old}}{\Omega_{new}}}$.

These factors multiply for a longer chain, so if we have larger and larger chains, our estimations for $J$ become exponentially mistaken. This is somewhat similar to the discrepancy of Eq. (\ref{eqn:betaJtildeDisc}) -- having "high order terms" in the Hamiltonian complicates following the flow.

This effect means if we have long spin chains, the numerical simulations will result with too low values of $\beta$. Since the numerics show a spin-dominant phase only for vanishing $J$ ($g_0\to 0$) either way, making them even smaller shouldn't change much; it might, however, affect the conductance of the system in this (insulating) phase and strengthen the claim that such phase will be insulating.

\section{Numerical Tests for the Validity of the Ansatz}
\label{app:Ansatz}
Throughout the text we have assumed that the Ansatz of subsection \ref{subsec:Ansatz} holds throughout the "blue" phases of Fig. \ref{fig:heatmap}, the superconductor and the Bose glass. We further deduced from this behavior to describe the expected behavior in the Spin Chain phase.

This appendix has two objectives. The first is to numerically test this Ansatz and its assumptions, justifying their usage. We are specifically interested in testing the validity of Eqs. (\ref{eqn:betaJtilde}) and (\ref{eqn:fzetanb}), regarding the behavior of D sites and B sites compared to R sites.

The second objective is to show what happens when the assumptions fail, at least partially. We shed light on the Spin Glass-Bose Glass transition with numerical graphs that complete the short discussion of subsection \ref{subsec:SC-BGTrans}.

We will start by discussing the charge sector, with $f(\zeta, \bar{n})$; then move on to the phase sector, with $g(\beta)$ and $h(\kappa)$; and finally discuss the deviations from the ansatz that can be observed in similar tools.

\subsection{Distributions of $\zeta$ and $\bar{n}$}

In Eqs. (\ref{eqn:fzetan}) and (\ref{eqn:fzetanb}) we implicitly assume that the distributions of $(\zeta, \bar{n})$ for R sites and B sites are strongly coupled -- specifically, that they share the same $f_0$. Apart from that, we also follow the single chain case\cite{PhysRevB.81.174528} and assume the distribution matches a specific structure; we shall not directly discuss this assumption here, but will see some of its implications.

The major implication of the ansatz regarding $\zeta$ and $\bar{n}$ is the special behavior for $f_0\to \infty$. If $f_0 \to 0$, the expected behavior is effectively

\begin{align}
    f(\zeta,\bar{n}) \approx f_0 e^{-f_0 \zeta},
\end{align}
and the restriction $\zeta>\zeta_c(\bar{n})$ can be replaced with $\zeta>0$, a simpler case where the approximation was previously shown to be justified.\cite{PhysRevLett.93.150402}

Our main tools to test this behavior will be the marginal distribution of $\bar{n}$ and the overall rates $\gamma_f,\gamma_f^b$. If $f_0\to \infty$, the distributions are expected to concentrate around the special values $\bar{n} =\pm \frac{1}{2(1+v)}$ (for R sites) and $\bar{n} = \pm \frac{1}{2}$ (for B sites). Observing this behavior will show that $\bar{n}$ follows the Ansatz even though its initial distribution was narrow, and had a low $\langle|\bar{n}|\rangle$. 

As RG time flows we approach the ansatz, and the exoected average is $\left\langle|\bar{n}|\right\rangle_B=\frac{1}{2\left(1-e^{-f_0}\right)}-\frac{1}{2f_0}$ and similarly for R sites. We conclude that for $f_0\to 0$ the average is $\frac{1}{4}$, and for $f_0\to \infty$ it approaches $\frac{1}{2}$.
\begin{figure}
\includegraphics[width=0.5\textwidth]{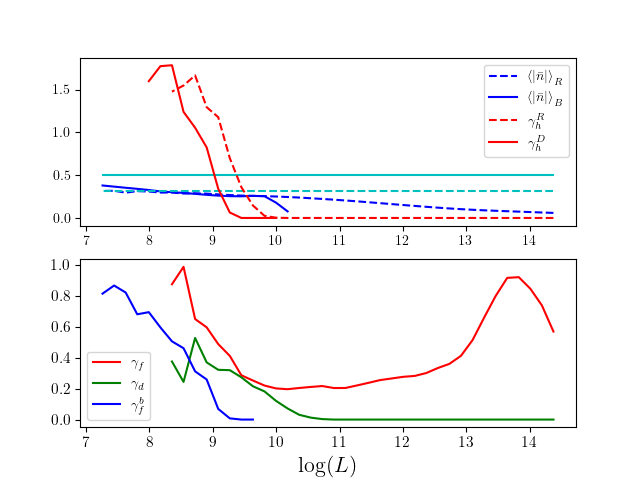}
\caption{
\label{fig:AppInsB}
Numerically extracted parameters for the case of a Bose Glass vs. the ladder length on a logarithmic scale, for $f_0=2.0, g_0=1.85$ and otherwise parameters matching Fig. \ref{fig:heatmap}. RG time flows \textbf{to the left}.\\
Top: $\left\langle\left|\bar{n}\right|\right\rangle$, calculated for R sites (blue, dashed) and B sites (blue, solid), compared to the relevant reference values (cyan); $\gamma_h$ calculated for R sites (red, dashed) and D sites (red, solid). \\
Bottom: $\gamma_f, \gamma_f^b$ and $\gamma_d$.
}
\end{figure}

The top plot of Fig. \ref{fig:AppInsB} shows this exact behavior. $\left\langle\left|\bar{n}\right|\right\rangle$ starts low and goes up monotonically, up to $\frac{1}{4}$ and then further up as $f_0$ rises. The full and dashed lines, representing B and R sites (correspondingly), stay together -- until $f_0\gg1$, where $\left\langle\left|\bar{n}\right|\right\rangle_R\approx \frac{1}{2(1+v)}\approx0.3$ and $\left\langle\left|\bar{n}\right|\right\rangle_B\approx \frac{1}{2}$.

It further shows that $\gamma_f$ and $\gamma_f^b$ follow a similar trend (though with a little delay) until all the R sites vanish and $\gamma_f$ cannot be estimated. Further, $\gamma_f^b$ rises up to $1$ as $f_0\to\infty$, matching Eq. (\ref{eqn:gammafb}), indicating both distributions indeed follow the distributions (\ref{eqn:fzetan}) and (\ref{eqn:fzetanb}) with the same $f_0$.

\subsection{Distributions of $\beta$ and $\gamma_h$}

The behavior of $\beta$ for R and B sites is identical -- $\beta$ is additive, following Eq. (\ref{eqn:betastep}). For D sites, we assume that $\tilde{J}$ can be parametrized with the same $\beta$, based on the relation from Eq. (\ref{eqn:betaJtilde}).

The distributions of $\beta_J$ and $\beta_{\tilde{J}}$, as discussed immediately after Eq. (\ref{eqn:betaJtilde}), could have been completely independent. They are connected by processes like the one captured in Eq. (\ref{eqn:UdecimJD}), and through the analysis we therefore assume that they are virtually identical.

To justify this assumption, we take a system very close to the Spin Glass-Bose Glass transition and observe different distributions. We calculate $\log\left(\frac{\Omega}{J}\right)$ and $\log\left(\frac{\Omega}{\tilde{J}}\right)$ for every type of site and expect to see $g_0^{-1}$ and $2g_0^{-1}$, depending on the type.

Selecting a typical set of parameters, we get these numerical results:

\begin{tabular}{|c|c|c|c|c|c|c|}
\hline
Connection Type & R-R & R-B & B-B & R-D & B-D & D-D \\ \hline
$\left\langle\log\left(\frac{\Omega}{J}\right)\right\rangle$ & 13.4 & 14.0 & 14.8 & 26.4 & 26.1 & 24.3 \\ \hline
\end{tabular}

and we see that the average $\beta$ is approximately uniform -- $\frac{\Omega}{\tilde{J}} \approx \left(\frac{\Omega}{J}\right)^2$; differences of approximately $\ln(2)\approx 0.7$ rise due to reasons similar to those described in the end of \ref{app:Bonds}.

As to $\kappa$, we have used the constant $\gamma_h$ to describe the behavior of $h(\kappa)$ (it is further discussed in the following subsection). In Fig. \ref{fig:AppInsB} we see $\gamma_h$ numerically estimated for R (red, dashed) and D (red, solid) sites. We see that it abruptly changes from $0$ to $\geq 1$, when the "zoom in" and addition processes (discussed around Fig. \ref{fig:kappa}) get to $\left|\kappa\right|$ of order unity. A similar graph for the spin glass phase is shown in Fig. \ref{fig:AppInsD}.

In both cases, we see that D sites lag behind R sites -- as they were generated from "old" R sites, with slightly smaller $\beta$ and $\kappa$ compared to the "present" R sites. We also see that the overall structure is very similar, hence approximating both distributions as equal captures the main features of the flow.

\subsection{$\gamma_f^b$ and $\gamma_h$ -- invalidity of the ansatz}

\begin{figure}
\includegraphics[width=0.5\textwidth]{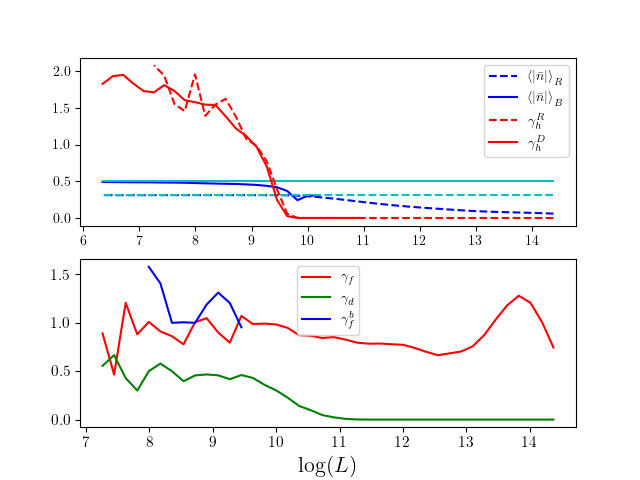}
\caption{
\label{fig:AppInsD}
Numerically extracted parameters for the case of a Spin Glass vs. the ladder length on a logarithmic scale, for $f_0=2.0, g_0=1.15$ and otherwise parameters matching Fig. \ref{fig:heatmap}. RG time flows \textbf{to the left}.\\
Top: $\left\langle\left|\bar{n}\right|\right\rangle$, calculated for R sites (blue, dashed) and B sites (blue, solid), compared to the relevant reference values (cyan); $\gamma_h$ calculated for R sites (red, dashed) and D sites (red, solid). \\
Bottom: $\gamma_f, \gamma_f^b$ and $\gamma_d$.
}
\end{figure}

We turn to try and understand the details of the Spin Glass-Bose Glass transition.

The prediction of the Ansatz is that for $f_0\to\infty$ and $g_0\to 0$ we shall have $\gamma_f^b \approx 1$ and $\gamma_h \approx 2$, hence according to Eq. (\ref{eqn:dsdgamma}) the only stable fixed point is $s=0$ and $b=1$.

A Spin Glass phase therefore can rise due to two main reasons: $\gamma_f^b, \gamma_h$ values that contradict the ansatz or initial $b=0$.

The numerical data shows that both effects take crucial parts. For the former reason we check Figs. \ref{fig:AppInsB} and \ref{fig:AppInsD} and observe $\gamma_f^b$ that reaches $1$ and slightly above; they also show that $\gamma_h$ does not reach $2$ until the system is very small. Similarly, in Fig. \ref{fig:AppInsB} we see $\left\langle|\bar{n}|\right\rangle_B$ that does not reach $\frac{1}{2}$, showing the Ansatz does not hold. We conclude that Eq. (\ref{eqn:fzetanb}) with $f_0 \to \infty$ is not a good model for the actual behavior of $f$.

These deviations are related, and can be explained with narrower $\bar{n}$ and $\kappa$ distributions. Low $|\bar{n}|$ explains the higher $\gamma_f^b$, as sites with lower $|\bar{n}|$ get decimated faster and low $|\kappa|$ directly explains a lower $\gamma_h$. We conclude that $\gamma_f^b=1$ and $\gamma_h=2$ are not very good descriptions of the system.

Simultaneously, as $g_0\to0$, the number of generated B sites gets smaller and smaller. The smaller $\gamma_g$ causes smaller $K$ values (as Eq. (\ref{eqn:kappastep}) is not applied as often) and more D sites are generated. The number of B sites is suppressed, and along with the larger $\gamma_h$ and smaller $\gamma_f^b$ we get a spin-dominated phase.

This analysis completes the discussion from subsection \ref{subsec:SC-BGTrans} and shows how the deviations from the ansatz described there affect the flow. Along with the Bose Glass-Superconductor transition, this completes our description of the different phases of the system and the basic nature of the phase transitions.

\bibliography{Bibliography}

\end{document}